%% file: main.tex
\documentclass[amssymb,amsmath,prd,twocolumn,preprintnumbers,superscriptaddress,nofootinbib]{revtex4-1}

\usepackage{graphicx}
\usepackage{bm}
\usepackage{amssymb,amsmath}
\usepackage{mathrsfs}
\usepackage{latexsym}
\usepackage{color}
\usepackage[normalem]{ulem} 
\usepackage{dcolumn}
\usepackage[colorlinks=true,citecolor=blue,urlcolor=blue]{hyperref}
\usepackage[usenames,dvipsnames]{xcolor}
\usepackage{xspace}
\usepackage{graphicx}

\usepackage{multirow}
\usepackage{xfp} %%% for basic arithmetic within tex
\usepackage{comment}
\usepackage{nameref}

\newcommand{\Rtyp}{\ensuremath{R_{1.4}}}

\newcommand{\rhonuc}{\rho_{\mathrm{nuc}}}

\newcommand{\Msolar}{\ensuremath{\mathrm{M}_\odot}}

\newcommand{\Mmax}{\ensuremath{M_{\rm max}}}

\newcommand{\veps}{\varepsilon}

\newcommand{\half}{\frac{1}{2}}

\DeclareMathOperator{\nuc}{nuc}

\usepackage[english]{babel}
\usepackage{xspace}
\usepackage{color,soul}

\newcommand{\quant}[1]{\left(#1\right)}

\include{macros}

\newcommand{\CIT}{\affiliation{Department of Physics, California Institute of Technology, Pasadena, California 91125, USA}}
\newcommand{\CITLab}{\affiliation{LIGO Laboratory, California Institute of Technology, Pasadena, California 91125, USA}}
\newcommand{\PI}{\affiliation{Perimeter Institute for Theoretical Physics, Waterloo, Ontario N2L 2Y5, Canada}}
\newcommand{\CITA}{\affiliation{Canadian Institute for Theoretical Astrophysics, University of Toronto, Toronto, Ontario M5S 3H8, Canada}}

\begin{document}

\title{
Implicit correlations within  phenomenological parametric models of the neutron star equation of state \\
}

\author{Isaac Legred}\CIT \CITLab
\author{Katerina Chatziioannou} \CIT \CITLab 
\author{Reed Essick} \PI
\author{Philippe Landry} \CITA

\begin{abstract}
    The rapid increase in the number and precision of astrophysical probes of neutron stars in recent years allows for the inference of their equation of state.
    Observations target different macroscopic properties of neutron stars which vary from star to star, such as mass and radius, but the equation of state allows for a common description of all neutron stars.  
    To connect these observations and infer the properties of dense matter and neutron stars simultaneously, models for the equation of state are introduced.  \emph{Parametric models} rely on carefully engineered functional forms that reproduce a large array of realistic equations of state.
    Such models benefit from their simplicity but are limited because any finite-parameter model cannot accurately approximate all possible equations of state.
    \emph{Nonparametric methods} overcome this by increasing model freedom at the cost of increased complexity.
    In this study, we compare common parametric and nonparametric models, quantify the limitations of the former, and study the impact of modeling on our current understanding of high-density physics.
    We show that parametric models impose strongly model-dependent, and sometimes opaque, correlations between density scales.
    Such interdensity correlations result in tighter constraints that are unsupported by data and can lead to biased inference of the equation of state and of individual neutron star properties.  
\end{abstract}

\maketitle

 %%%%%%%%%%%%%%%%%%%%%%%%%%%%%%%%%%%%%
\section{Introduction}
 %%%%%%%%%%%%%%%%%%%%%%%%%%%%%%%%%%%%%
 
 The equation of state (EoS) of the dense matter inside neutron stars (NSs) is uncertain at densities near and beyond nuclear saturation, $\rhonuc = 2.8 \times 10^{14} \mathrm{g/cm^3}$, because it cannot be precisely constrained by theoretical calculations or terrestrial experiments \cite{Lattimer:2015nhk,Ozel:2016oaf, Oertel:2016bki,Baym:2017whm, Machleidt:2011zz, Han:2019bub, Chatziioannou:2020pqz}.
 Astronomical observations \cite{TheLIGOScientific:2017qsa, Cromartie:2019kug, Miller:2019nzo, Riley:2019yda, Miller:2021qha, Riley:2021pdl,Antoniadis:2016hxz, LIGOScientific:2020aai, Raaijmakers:2019qny} target the macroscopic properties of NSs, such as their mass $M$, radius $R$, and dimensionless tidal deformability $\Lambda$, which in turn can be used to constrain the EoS at densities greater than nuclear saturation~\cite{Lattimer:2000nx,Lindblom:1992, Lindblom:2010bb}.

 A set of observations of different systems can be used to constrain a shared underlying property through a hierarchical inference scheme.
 The hierarchical formalism is derived in the context of combining data from different sources while faithfully incorporating their uncertainties and potential observational selection effects \cite{Loredo:2004nn}; see, e.g., Refs.~\cite{Chatziioannou:2020pqz, Landry:2020vaw}.
 In the context of NS structure, the main objective is to obtain a posterior for the EoS as a shared variable among many astrophysical observations.
 The prior corresponding to this posterior is not necessarily straightforward to define because the space of potential EoS (i.e., the space of possible functions obeying basic physical constraints) that relate the pressure $p$ and the baryon density $\rho$, $p = p(\rho)$, is infinite dimensional.\footnote{We can equivalently use  $p(\varepsilon)$, with $\varepsilon$ the internal energy density.  In the zero-temperature limit, ${d\varepsilon}/{d\rho} = \quant{p(\rho)+\varepsilon(\rho)}/{\rho}$.}

The simplest way to define such a prior is through a \emph{parametrization} of the EoS, which is a functional form of $p(\rho)$ that typically depends on a few parameters.
Common phenomenological models such as piecewise-polytrope~\cite{Read:2008iy},  spectral~\cite{Lindblom:2010bb, Lindblom:2013kra} and speed-of-sound~\cite{Greif:2018njt} parametrizations have been used to effectively sample candidate EoS for use in inference.
The simplicity of a closed-form parametric expression comes at the cost, though, of being unable to faithfully represent many of the possible degrees of freedom in the true EoS.
While many of these models can accurately represent most EoS derived from effective nuclear interactions \cite{Lindblom:2010bb, Read:2008iy}, it is not always clear how to extend these parametrizations toward more general behavior in the EoS that may arise from phase transitions or new physics. This limitation of phenomenological parametric models has been recognized from the outset~\cite{Read:2008iy}. However, in this study we investigate another way that they may artificially restrict the inferred EoS. %These limitations of phenomenological parametric models have been recognized even during their design~\cite{Read:2008iy}, however in this study we investigate a perhaps even more worrisome feature of theirs.

Parametric models use only a few parameters, which means that the values of $p(\rho)$ at different densities are often \emph{correlated}.
These correlations represent a source of \emph{model dependence} in the inference, which is undesirable insofar as it does not reflect true prior knowledge of the EoS at those densities. %because we do not have sufficient information about the EoS at these densities to justify such modeling assumptions. 
That is, the correlations induced by the choice of parametrization can constitute strong, unintentional prior beliefs about the EoS. This unwanted model dependence is a natural consequence of the phenomenological nature of the parametric models. %Such unwanted model dependence is not unexpected, as the parametric models are phenomenological and often enforce certain properties in the EoS.

An alternative method for constructing a prior on the space of EoS, which we call \emph{nonparametric} in what follows, targets more model flexibility by making use of Gaussian processes (GPs)~\cite{Landry:2018prl}.
This approach produces a multivariate Gaussian distribution for the function $\phi =  \log\quant{(c/c_s)^{2} - 1}$, where $c_s$ is the speed of sound and $c$ is the speed of light. 
By conditioning the prior only weakly on existing nuclear-theory models, we generate a \emph{model-agnostic} prior process for EoS. 
The chosen correlations between $\phi(p_i)$ and $\phi(p_j)$, or equivalently between the values of the EoS at different densities, are set by a \emph{kernel} function, which is in turn described by a few parameters.
%Model freedom, in this case, is controlled by chosen correlations between the values $\phi(p_i)$ and $\phi(p_j)$, equivalently the value of the EoS at different densities,  which is set by a \emph{kernel} function.
%The kernel function is, in turn, described by a few parameters.
Following Ref.~\cite{Essick:2019ldf}, we consider a variety of possible kernel parameters to probe a range of different correlations and thus maximize model freedom.
This approach allows us, in principle, to model any function $p(\rho)$, and furthermore to probe a wide range of interdensity correlations and high-density EoS behavior.

Of course, completely unrestricted freedom in the EoS is neither desirable nor realistic, as certain physical constraints should be encoded into the EoS prior. %Of course, unrestricted model freedom might not always be desirable or realistic as certain physical constraints should be encoded into the prior for the EoS. 
For example, an EoS must be causal, 
\begin{equation}
    \label{causality}
\frac{dp}{d\varepsilon} = c_s^2 < c^2,
\end{equation}
%k
and thermodynamically stable, 
\begin{equation}
    \label{stability}
\frac{dp}{d\varepsilon} = c_s^2 > 0.
\end{equation}
Imposing these constraints in the prior is desirable as it excludes unphysical models from the analysis.\footnote{Some analyses allow the EoS to be slightly acausal at times; see Appendix~\ref{parametric-eos-appendix} for more discussion.}

In this paper, we examine common parametric and nonparametric EoS models to determine the extent to which each prior's assumptions impact inference of the EoS and NS properties.
We find that the three parametric models we study (spectral, piecewise polytrope, and speed of sound) build additional interdensity correlations into the EoS beyond what can be attributed to causality and stability.
These correlations between densities typically lead to more stringent constraints than are strictly supported by the data.
On the other hand, the nonparametric model demonstrates the largest degree of model independence, restricted primarily only by causality and thermodynamic stability. 
We demonstrate that these strong, model-dependent interdensity correlations have already impacted inferred microscopic and macroscopic NS properties. 
Such effects are expected to become more severe as statistical uncertainties decrease with more data that probe different NS densities.

The remainder of the paper is organized as follows.
In Sec.~\ref{methods-and-motivation}, we describe our inference methods and our approach to investigating model dependence. %motivate our investigation into model dependence.
In Sec.~\ref{impact-current-constraints}, we examine the EoS and NS properties inferred with current data and show that they are influenced by correlations in the EoS prior.
In Sec.~\ref{sec:toy model}, we illustrate the main limitations of parametric EoS inference with a toy model.
In Sec.~\ref{interdensity-correlations}, we quantify the implicit EoS correlations and demonstrate that the nonparametric model displays the largest degree of model independence.
In Sec.~\ref{mock-astro}, we study the correlations' potential impact on upcoming EoS inference using mock astrophysical measurements.
In Sec.~\ref{parametric-prior-choices}, we demonstrate that the limitations identified in parametric models cannot be resolved by making small modifications to the prior distributions.
Finally, in Sec.~\ref{discussion} we discuss our conclusions.  

 %%%%%%%%%%%%%%%%%%%%%%%%%%%%%%%%%%%%%
\section{Methods and Models}%Motivation}
\label{methods-and-motivation}
 %%%%%%%%%%%%%%%%%%%%%%%%%%%%%%%%%%%%%

The posterior for the EoS depends on two elements: (i) the prior EoS process, and (ii) the data.
Our goal in this study is to assess the effect of the prior as generated from different parametric and nonparametric models for the EoS.
We therefore always employ the same data, which we briefly describe in Sec.~\ref{data}. The hierarchical likelihood corresponding to this data is described in Refs.~\cite{Landry:2020vaw, Legred:2021}. 
The EoS priors are described in detail in Sec.~\ref{priorprocess}, where we discuss the different EoS models and parameter priors that generate each EoS prior process. 

%%%%%%%%%%%%%%%%%%%%%%%%%%%%%%%%%%%%%
\subsection{EoS prior}
\label{priorprocess}

We wish to establish a prior process over candidate EoS. By this, we mean a probabilistic measure on the space of potential EoS.  
To do this, we use several \emph{models} of the EoS. We distinguish \emph{parametric} models,  which provide a functional form for the EoS, from nonparametric models which do not impose such a functional form.   
We use three different phenomenological parametric models, a \emph{piecewise-polytrope}~\cite{Read:2008iy} parametrization, a \emph{spectral} parametrization~\cite{Lindblom:2010bb}, and a direct parametrization of the \emph{speed of sound}~\cite{Greif:2018njt}.
The spectral and piecewise-polytrope parametrizations use a polytropic form for the EoS, so that
$$
    p(\rho) = K\rho^\Gamma.
$$
In the piecewise-polytrope case, the polytropic index $\Gamma$ is a piecewise-constant function of the pressure, while in the spectral case, $\log (\Gamma)$ is expanded as a polynomial in pressure.
In the speed of sound parametrization, the speed of sound is expressed as a constant plus a Gaussian and a logistic curve which asymptotes to $c^2/3$. Following past practice~\cite{Carney:2018sdv}, we slightly relax the causality threshold and consider EoS with $c_s^2<1.1c^2$ for all parametric models.
See Appendix~\ref{parametric-eos-appendix} for more details about each model and its implementation.

To establish a prior process, we must additionally supply a joint prior probability distribution on the parameters of each model from which a \emph{draw} is a realization of the parameters and therefore a candidate EoS.
For example, in the spectral model, the parameters are coefficients in the spectral expansion.
In the piecewise polytrope, the parameters are the value of the polytropic index itself.
For our headline results, we use standard priors for the parametric models~\cite{Carney:2018sdv, Wysocki:2020myz}, except for the speed-of-sound model, which we adapt to increase access to astrophysically relevant EoS; again, see Appendix~\ref{parametric-eos-appendix} for details.  

We compare the prior processes generated by the parametric models to a prior process from a nonparametric model~\cite{Landry:2018prl}.
While our nonparametric implementation does not assume a specific functional form for the EoS, it does parametrize the correlations between the sound speed at different densities. These correlations are described by a kernel function.
In practice, we choose a large set of points, $p_i$, and then the variable $\phi(p_i)$ is sampled from a multivariate Gaussian distribution.
By changing the kernel's parameters and conditioning on different nuclear models, we can generate a range of GPs.
We choose a model-agnostic prior, which is to say we average over multiple GPs with different correlations, each loosely informed by nuclear-theory models~\cite{Landry:2018prl}.
We do this to maximize the freedom of the model.
See Appendix~\ref{np-appendix} for more details.  

Due to its construction, the GP itself has parameters which control correlations.
Such parameters have been termed \emph{hyperparameters} \cite{Landry:2018prl}, though we avoid this terminology here in order to avoid potential confusion with the term's use in hierarchical inference.
In addition, the parametric models also have parameters which control the prior process; in general, such details are unique to each model.
We instead focus primarily on the prior process induced by each EoS model with its chosen prior, returning to the subject of parameter distributions briefly in Sec. \ref{parametric-prior-choices}.
For now, we simply note that our model implementations are  typical of those used in the literature~\cite{Read:2009yp, Lindblom:2010bb, Carney:2018sdv, Lackey:2014fwa, Greif:2018njt}.  
Lastly, we stress that our distinction between parametric and nonparametric models lies not in the existence of parameters but in the specification of a functional form for the EoS.  In particular, we compare models with small, fixed numbers of parameters that are commonly used in the literature. For these models, the choice of functional form significantly impacts the range of EoS that can be represented.

%-------------------------------------
\subsection{Data and likelihood}
\label{data}

Unless otherwise stated, all analyses in this paper make use of the same astronomical data as Ref.~\cite{Legred:2021}.
Specifically, we include two mass-tidal deformability measurements from gravitational wave (GW) detections of merging NSs~\cite{Abbott:2018exr, Abbott:2020uma}, one heavy pulsar mass measurement with radio data~\cite{Antoniadis:2013pzd}, and two x-ray observations of NS masses and radii~\cite{Miller:2019nzo, Riley:2019yda, Miller:2021qha, Riley:2021pdl}.
In the latter case, we also use the up-to-date radio mass measurement of the pulsar J0740+6620~\cite{Fonseca:2021wxt}. 
Given these data $d$, the posterior probability density of a particular EoS $\veps$ is  
\begin{equation}
  P(\veps|d,\mathcal I) = \frac{P(d|\veps, \mathcal I)}{P(d|\mathcal I)}P(\veps |\mathcal I),
\end{equation}
where $\mathcal I$ is any additional information we may have about the system, e.g., knowledge that the data originate from a NS, as is the case for the pulsar observations but not for the GWs.
Here $P(\veps|d,\mathcal I)$ is the \emph{posterior} probability of the EoS given the data, $P(d|\veps, \mathcal I)$ is the \emph{likelihood} of the astrophysical data given the EoS, $P(\veps|\mathcal I) $ is the prior probability of the EoS, and $P(d|\mathcal I) = \int  P(d|\veps, \mathcal I) P(\veps|\mathcal I) \mathcal{D}\veps $ is the total probability of observing this data marginalized over all  EoS in the prior, often called the \emph{evidence}.
For general astrophysical data, $P(d|\veps, \mathcal I)$ must be computed by marginalizing over the astrophysical distribution of masses, spins, sky locations, and distances for individual events, which remains poorly constrained~\cite{Alsing:2017bbc,FarrChatziioannou2020,Chatziioannou:2020msi,Fishbach:2020ryj,LIGOScientific:2021psn,Landry:2021hvl,Farah:2021qom}.
For the full expression, see Refs.~\cite{Landry:2020vaw,Chatziioannou:2020pqz}.

The different datasets we use primarily inform the EoS at different densities.
The heaviest pulsar mass measurements serve to downweight EoS which cannot support the observed NS masses; these constraints tend to most significantly impact inference near $\sim (4$-$6)\rhonuc$ and typically favor a stiffer EoS.
The x-ray data provide constraints on the NS radius, and constraints so far have given information about the EoS mainly in the region $\sim (1$-$4)\rhonuc$~\cite{Miller:2021qha, Landry:2020vaw, Legred:2021, Raaijmakers:2021uju}.
The GW observations provide constraints on the tidal deformabilities of the binary components, which are dominated by the loudest event observed so far, GW170817~\cite{TheLIGOScientific:2017qsa, Landry:2020vaw}. 
In terms of densities, the relevant scale constrained by this measurement is $\sim (1$-$3)\rhonuc$~\cite{Landry:2020vaw, Legred:2021}.
Future constraints with GWs are likely to lie in this density range, as the fractional uncertainty in $\Lambda$ will be smallest for lower-mass NSs with less dense cores and larger tidal deformabilities.
In principle, nuclear experiments or calculations could also be included in such an analysis, and would mainly constrain the EoS near or below $\rhonuc$~\cite{Essick:2021ezp, Essick:2021kjb, Pang:2021jta, Biswas:2021yge, Raaijmakers:2021uju}.
However, we do not incorporate any in this work.

 %%%%%%%%%%%%%%%%%%%%%%%%%%%%%%%%%%%%%
 \section{Impact of EoS model on current EoS constraints}
\label{impact-current-constraints}
 %%%%%%%%%%%%%%%%%%%%%%%%%%%%%%%%%%%%%

%
\begin{figure}
\centering
    \includegraphics[width=.49\textwidth]{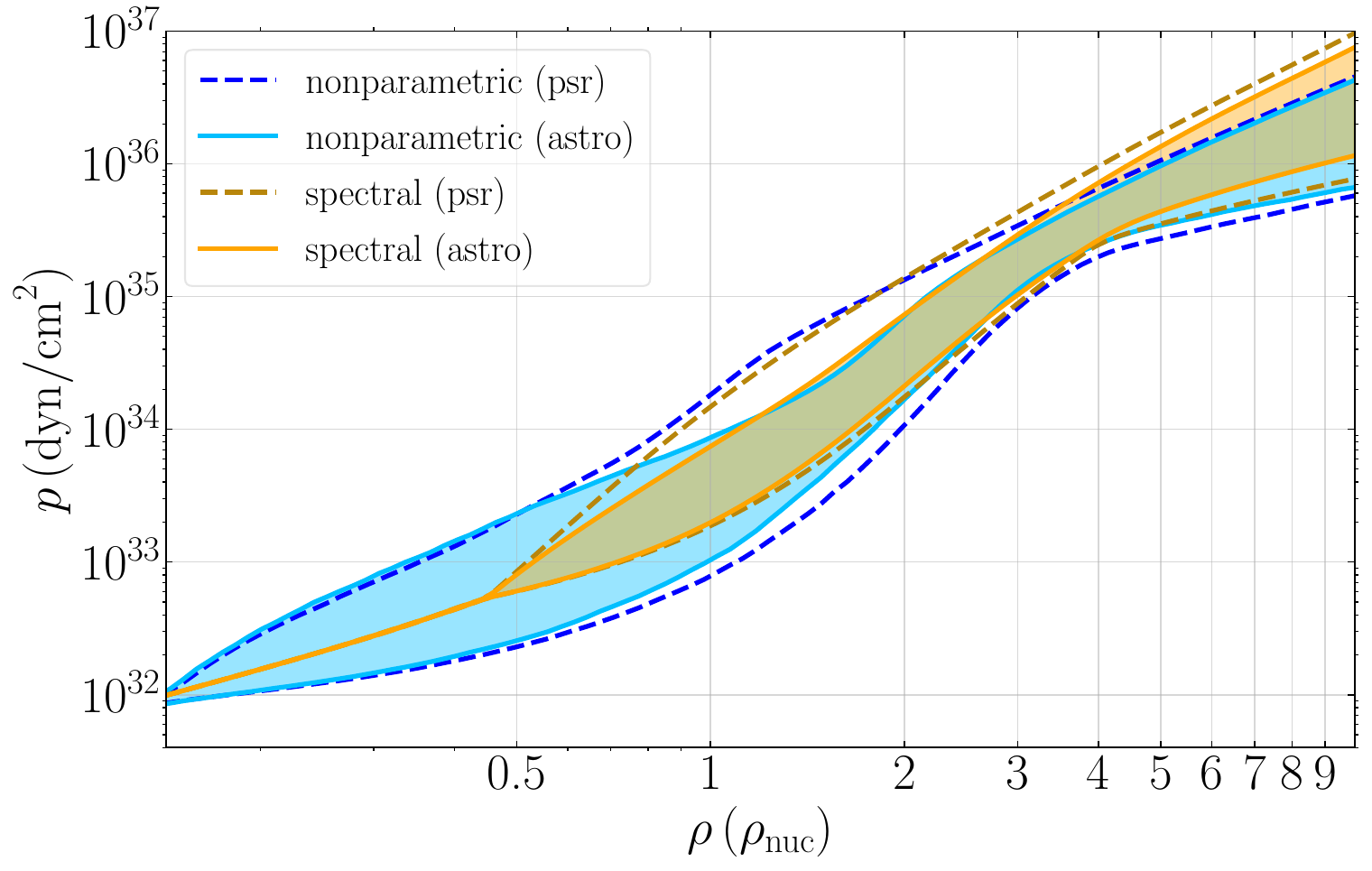}
    \caption{
        Symmetric 90\% credible region for the pressure $p$ at each density $\rho$ in units of the nuclear saturation density using the nonparametric and spectral prior processes.
        We show results including all astrophysical data (labeled ``astro," solid lines) and restricting to the heavy pulsars only (labeled ``psr," dashed lines).
        The latter choice ensures that prior choices on the $\Mmax$ supported by each model are irrelevant.
        Other parametric models are shown in Fig.~\ref{fig:prho_mr_all_models}.
        In all cases, we find that the $p$-$\rho$ posterior depends on the EoS model even when identical data and inference schemes are employed. 
    }
    \label{fig:process_prior_plot}
\end{figure}
\begin{figure*}
  \centering
  \includegraphics[width=.49\textwidth]{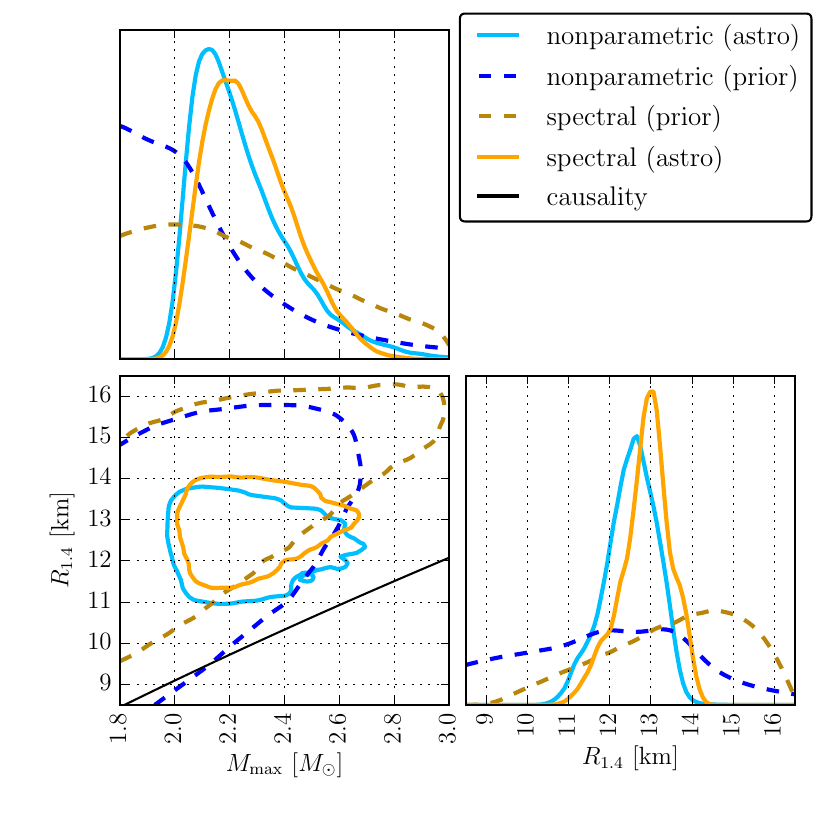}
  \includegraphics[width=.49\textwidth]{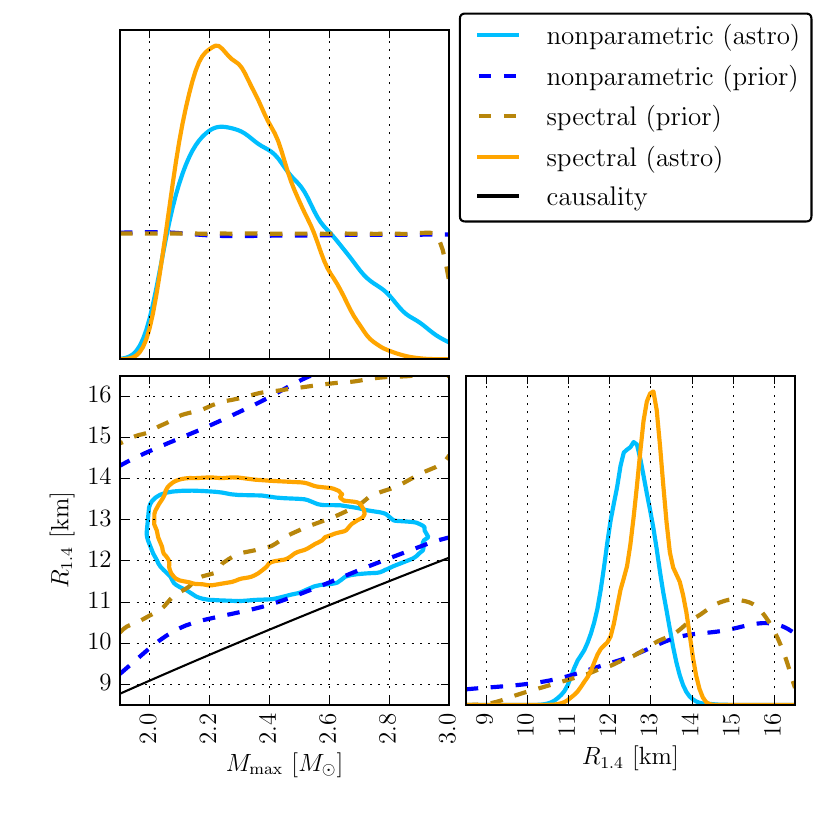}
    \caption{
        Prior (dashed) and posterior (solid) for the radius of a 1.4$M_{\odot}$ NS, $\Rtyp$, and maximum mass, $\Mmax$, of a NS for two choices of the marginal $\Mmax$ prior: default (left) and flat (right).
        We show results with the nonparametric and spectral EoS models, and contours denote  90\% credible regions.
        The black line in the two-dimensional plot represents a maximally stiff $M$-$R$ curve~\cite{Rhoades:1974fn}  stitched to a fiducial low density EoS~\cite{Kalogera:1996ci}.
        Due to the low-density stitching, this ``causality" line should be interpreted as a fuzzy boundary and not a sharp line.
        Both panels demonstrate that the spectral and nonparametric EoS models produce different $\Mmax$ posteriors and that these differences cannot be attributed to the marginal priors.
        They are instead caused by correlations between low and high densities (equivalently, between $\Mmax$ and $\Rtyp$) imposed by the models.
        The correlations in the nonparametric case are due to causality, while the spectral case exhibits additional correlations and model dependence.
    }
\label{fig:spectral_Mmax-Rtyp}
\end{figure*}

Following the above prescription, we analyze the existing data using the four different EoS priors and plot the resulting marginal posteriors for $p(\rho)$ across a wide range of densities. Figure~\ref{fig:process_prior_plot} compares the spectral and nonparametric models; similar plots for the other parametric models can be found in Appendix~\ref{other-parametric-results}.
The posteriors differ in their predictions for the EoS.
For instance, the spectral posterior is stiffer on average than the nonparametric one, especially above $4\rhonuc$. Similar differences have also been pointed out in Refs.~\cite{Greif:2018njt,Miller:2021qha,Raaijmakers:2021uju}, where multiple EoS models were employed under identical analysis settings.
Our goal here is to understand the origin of these discrepancies.

%Since envelope plots are hard to interpret across different densities,
Since it is difficult to glean information about interdensity correlations from envelope plots like Fig.~\ref{fig:process_prior_plot}, we turn our attention to two macroscopic NS properties that roughly correspond to the EoS behavior at high and low densities: the maximum mass, $\Mmax$, and the radius of a 1.4$M_{\odot}$ NS, $\Rtyp$.
In Fig.~\ref{fig:spectral_Mmax-Rtyp} (left panel), we plot the one- and two-dimensional marginal prior and posterior for $\Mmax$ and $\Rtyp$.
As expected, the marginal posteriors differ, but so do the marginal priors.
Indeed, the $\Mmax$ plot shows that both the spectral prior and posterior seem to have more support for $\Mmax$ around $2.2-2.5M_{\odot}$ than the nonparametric case.
However, this trend is reversed above $2.6\,M_{\odot}$.
This observation suggests that the difference between the nonparametric and spectral posteriors cannot be trivially assigned to different marginal priors.
To further demonstrate this, in the right panel we plot the same variables, but now reweighted to a flat marginal $\Mmax$ prior.
As expected, the two posteriors differ, with the spectral model producing a narrower posterior.

To understand this discrepancy, we revisit the possible reasons $\Mmax$ is constrained on the high side.
Though an upper limit on $\Mmax$ has been proposed based on the analysis of the counterpart of GW170817~\cite{Margalit:2017dij}, our analysis does not make use of it. 
Excluding an origin due to data, the upper limit on $\Mmax$ must be the result of the EoS prior.
The two-dimensional $\Mmax$-$\Rtyp$ panel indeed shows that the upper limit on $\Mmax$ is related to the upper limit on $\Rtyp$~\cite{TheLIGOScientific:2017qsa, Abbott:2018exr}; the $\Mmax$-$\Rtyp$ prior does not cover the entire available region for either model, with larger $\Mmax$ requiring stiffer EoS and larger $\Rtyp$.

The fact that larger $\Mmax$ requires large values of $\Rtyp$ is not unexpected from causality considerations. 
Indeed, the causality condition and the pressure at twice saturation, $p_{2.0}$, set an upper limit on the value of pressure at five times saturation, $p_{5.0}$.  
Since $p_{2.0}$ and $p_{5.0}$ correlate with $\Rtyp$ and $\Mmax$, respectively~\cite{Lattimer:2000nx}, any causal EoS model should limit $\Mmax$ for certain low $\Rtyp$ configurations~\cite{Rhoades:1974fn}.
To quantify this, we overplot the limiting $\Mmax$-$\Rtyp$ relation given by Ref.~\cite{Kalogera:1996ci}: each point on the line represents a soft low-density EoS stitched to an EoS with $c_s^2 = 1$ at different densities.
This curve should be interpreted approximately, as the exact causality threshold depends on the details of the low-density EoS~\cite{Drischler:2021bup}.

Nonetheless, the right panel shows that in the nonparametric case the prior fills more of the physically allowable $\Mmax$-$\Rtyp$ parameter space compared to the spectral prior. This indicates that the nonparametric prior has non-negligible support in the entire physically allowed region even if specific marginal priors might downweight some regions (left panel). The same is not true for the spectral model which cannot access certain regions of the $\Mmax$-$\Rtyp$ plane.\footnote{Figure~\ref{fig:spectral_Mmax-Rtyp} shows 90\% contours. If we plotted 99\% contours instead, the nonparametric model accesses even more of the allowed space, while the spectral model remains restricted.  }
This demonstrates that the correlation between $\Mmax$-$\Rtyp$ that appears in the spectral model is not entirely due to causality considerations, but it is also affected by the specifics of the model.
The nonparametric model, however, is able to produce EoS which fall near the causality limit, indicating physics rather than modeling artifacts are the primary limitation to model freedom.
Figure~\ref{fig:piecewise-sos-MR} presents a qualitatively similar conclusion for the piecewise-polytrope and speed-of-sound models. The piecewise polytrope exhibits  behavior similar to the spectral model, while the speed-of-sound parametrization exhibits the opposite problem: its prior does not include low $\Mmax$ for large $\Rtyp$ values.

Crucially, the correlations we see in these corner plots are only those that are apparent from the two-dimensional marginalized posteriors. They do not reveal the many hidden correlations within the parametric EoS models that are not as easily detected.
It is possible for implicit correlations within the EoS prior to bias the inference in ways that are not obvious in low-dimensional projections.

%%%%%%%%%%%%%%%%%%%%%%%%%%%%%%%%%%%%%%
\section{Impact of interdensity correlations: Toy Model}% for EoS Inference}
\label{sec:toy model}
%%%%%%%%%%%%%%%%%%%%%%%%%%%%%%%%%%%%%%

To better understand the effect of such implicit correlations, we first consider a simple toy model that demonstrates several of the issues with parametric models and introduce our techniques for diagnosing them.

We consider several simple linear parametrizations of the pressure as a function of energy density $p(\varepsilon)$.
This allows us to examine the prior processes induced by the assumption of linearity from various perspectives.
We contrast this to a GP prior process in the same context, finding particularly striking differences in the effect of a precise measurement of the pressure at one density on our uncertainty in the pressure at other densities.

\begin{figure}
    \centering
    \includegraphics[width=1.0\columnwidth, clip=true, trim=0.0cm 0.4cm 0.0cm 0.0cm]{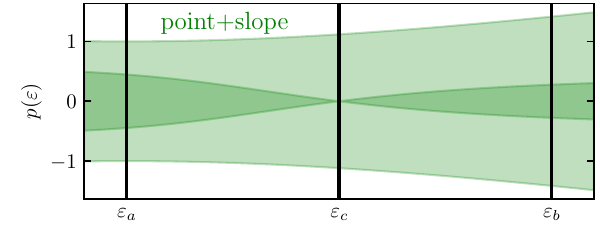} \\
    \includegraphics[width=1.0\columnwidth, clip=true, trim=0.0cm 0.4cm 0.0cm 0.0cm]{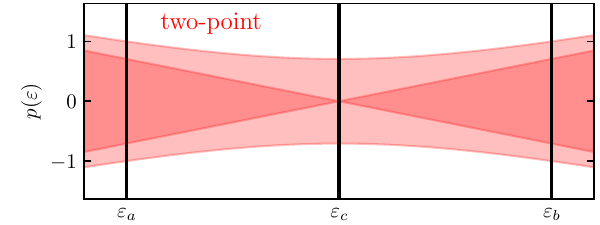} \\
    \includegraphics[width=1.0\columnwidth, clip=true, trim=0.0cm 0.0cm 0.0cm 0.0cm]{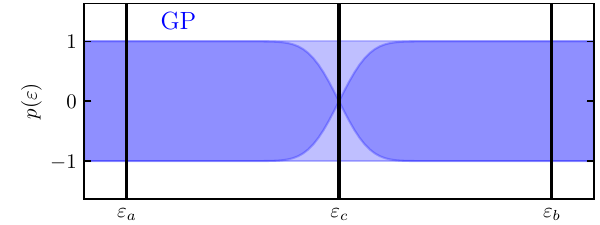}
    \caption{
        68\% ($\pm$ 1-$\sigma$) marginal credible regions for the pressure at each density under the point+slope (top, green), two-point (middle, red), and GP (bottom, blue) prior processes.
        Shaded regions correspond to marginal distributions induced by the prior process at each energy density (light colors) and conditioned distributions for $p(\varepsilon)$ given a precise observation of $p_c$ (dark colors).
        Only the GP process ``fills the prior volume'' rapidly as one moves away from the observation point $\varepsilon_c$.
        Compare to Fig.~\ref{fig:mock_pressure_injections_comparison}.
    }
    \label{fig:toy model envelope}
\end{figure}

We begin with the simple parametric model of a linear relationship between the pressure and the energy density
\begin{equation}
    p(\varepsilon) = p_a + c_s^2 (\varepsilon - \varepsilon_a),
\end{equation}
parametrized by the pressure at $p_a = p(\varepsilon_a)$ and the slope $c_s^2$.
Ignoring causality constraints, we choose what appear to be uninformative priors
\begin{gather}
    p_a \sim \mathcal{N}(\mu_a, \sigma_a^2), \quad c_s^2 \sim \mathcal{N}(\mu_{c_s^2}, \sigma_{c_s^2}^2),
\end{gather}
and refer to this as the point+slope process.
The top panel of Fig.~\ref{fig:toy model envelope} shows the envelope plot for this prior process, i.e., the marginal distributions of the pressure at each energy density.

\begin{figure}
    \centering
    \includegraphics[width=1.0\columnwidth, clip=True, trim=0.0cm 0.5cm 0.0cm 0.0cm]{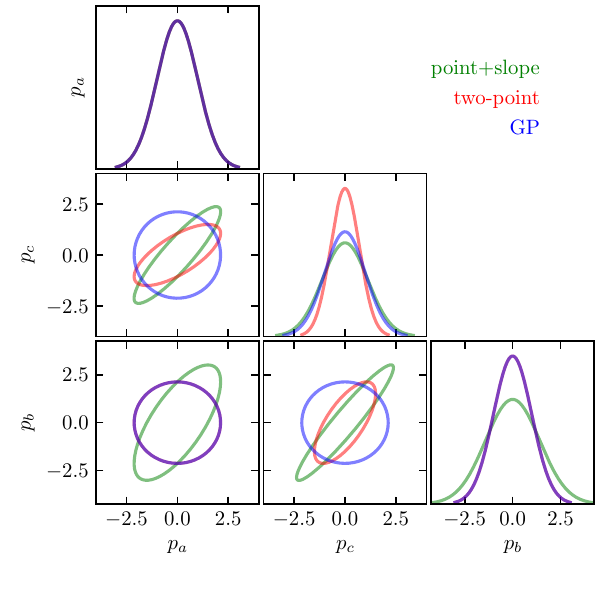}
    \caption{
        Joint and marginal distributions for the pressures at the three reference densities called out in Fig.~\ref{fig:toy model envelope} for the point+slope (green), two-point (red), and GP (blue) prior processes.
        Contours in the joint distribution represent 90\% credible regions. 
        While certain choices for the parametrization and priors can uncorrelate pairs of variables, only the GP prior process induces minimal correlations between all variables simultaneously.
    }
    \label{fig:toy model corner}
\end{figure}

The envelope plot appears reasonable.
That is, the prior process assigns approximately equal uncertainty to each pressure.
However, the envelope plot only shows the marginal distributions at each density.
Figure~\ref{fig:toy model corner} shows the correlations between pressures at different densities.
From this we see that the prior process actually imposes strong correlations between all pressures.
We quantify this correlation between $p_a$ and the pressure at some other density $ p_b \equiv p(\epsilon_b) $ with the \emph{mutual information}~\cite{InfoTheory}, defined as 
\begin{equation}
    I(p_a, p_b) \equiv \int dp_a dp_b \, P(p_a, p_b) \ln \left( \frac{P(p_a, p_b)}{P(p_a) P(p_b)} \right),
\end{equation}
where $P(p_a) = \int dp_b \, P(p_a, p_b)$ is the marginal distribution. For the point+slope parametrization, we compute
\begin{equation}
    I(p_a, p_b) = \frac{1}{2} \ln \left( 1 + \frac{\sigma_a^2}{\sigma_{c_s^2}^2 (\varepsilon_a - \varepsilon_b)^2} \right).
\end{equation}

The mutual information between $p_a$ and $p_b$ can be made arbitrarily small only in the limit $\sigma_a \ll \sigma_{c_s^2} |\varepsilon_b - \varepsilon_a|$; however, this limit corresponds to vanishingly small marginal uncertainty for $p_a$.
We conclude that the assumption of a linear functional form can produce what seems to be a reasonable envelope plot in Fig.~\ref{fig:toy model envelope}, but nevertheless induces model-dependent correlations between the pressure at different densities in Fig.~\ref{fig:toy model corner}.

In an attempt to remove  the correlation between $p_a$ and $p_b$, we consider the alternative parametrization 
\begin{equation}
    p(\varepsilon) = p_a + \frac{p_b - p_a}{\varepsilon_b - \varepsilon_a} (\varepsilon - \varepsilon_a),
\end{equation}
described by the pressures at the two reference densities.
We assume priors
\begin{gather}
    p_a \sim \mathcal{N}(\mu_a, \sigma_a^2), \quad p_b \sim \mathcal{N}(\mu_b, \sigma_b^2),
\end{gather}
and refer to this as the two-point process.
Figures~\ref{fig:toy model envelope} and~\ref{fig:toy model corner} show envelope and marginal distributions, respectively.

By construction, $I(p_a, p_b) = 0$ for the two-point prior process, as also seen in Fig.~\ref{fig:toy model corner}.
However, the envelope plot shows that the marginal prior actually tightens for pressures between the reference densities.
That is, we are able to remove the correlation between two pressures only at the expense of asserting greater prior knowledge about other pressures.
Additionally, Fig.~\ref{fig:toy model corner} shows that there are still correlations between ($p_a$, $p_b$) and other pressures.
This hints at the fact that, when one assumes a specific functional form, it may be possible to remove the correlations between a small number of statistics, but it is generally difficult to make all correlations vanish simultaneously or to avoid making strong assumptions about specific values of the function.

In order to consider this effect more quantitatively, we introduce a generalization of the mutual information that considers three pressures~\cite{InfoTheory}
\begin{align}
    &I(  p_a, p_b, p_c) \nonumber \\
        & \equiv \int dp_a dp_b dp_c \, P(p_a, p_b, p_c) \ln \left( \frac{P(p_a, p_b, p_c)}{P(p_a) P(p_b) P(p_c)} \right) \nonumber \\
        & = \int dp_a dp_b \, P(p_a, p_b) \int dp_c \, P(p_c|p_a, p_b) \ln \left( \frac{P(p_c|p_a, p_b)}{P(p_c)} \right) \nonumber \\
        & \quad\quad + I(p_a, p_b).\label{mutual_info_gen}
\end{align}
Even if one can choose parametrizations and priors such that $I(p_a, p_b)$ vanishes, there is another term when considering mutual information for three pressures.
In fact, for both the point+slope and two-point prior processes, the integral over $p_c$ diverges as $P(p_c|p_a, p_b)$ is a delta function (determined by the closed-form parametrization), while $P(p_c)$ is a Gaussian with finite width.
We conclude that the assumption of a linear relationship between the pressure and the density implies an infinite amount of information about the allowed relationships between variables.
One cannot undo all these correlations at the same time by a clever choice of marginal prior distributions, although it may be possible to undo some of them.
The failure of this reparametrization scheme anticipates the results of our investigation of alternative parametric priors in Sec.~\ref{parametric-prior-choices}.

In general, the only way to undo all correlations simultaneously is to add more model freedom into the prior process.
For example, one may add more reference densities to an existing model and generate a piecewise linear prior process.
However, there will always be some densities between the (finite number of) reference densities, regardless of how many reference densities are chosen.
In each of those regions, the piecewise-linear model is equivalent to our two-point prior process, and the strong correlations remain.
One is then left with the question of how to extend the parametrization to remove all correlations in a scalable way.
We show that a GP is a natural solution.

A GP, defined in terms of a mean function and a covariance kernel, describes our uncertainty in the infinitely many degrees of freedom in a function.
With the assumption of Gaussianity, we can easily marginalize away uninteresting degrees of freedom, in our case retaining only the pressures on a dense grid of energy densities, and the GP reduces to a high-dimensional multivariate Gaussian distribution.
Specifically, we consider the joint distribution induced over $p_a$, $p_b$, and $p_c$ by a GP
\begin{equation}
    \vec{p} \sim \mathcal{N}(\vec{\mu}, \Sigma),
\end{equation}
with mean $\vec{\mu}$ and covariance $\Sigma$, with matrix elements defined by a covariance kernel
%
% \begin{equation}
%     \vec{\mu} = (\mu_a, \mu_b, \mu_c),
% \end{equation}
% %
% and covariance
% %
% \begin{equation}
%     \Sigma = \begin{bmatrix} \sigma_{aa}^2 & \sigma_{ab}^2 & \sigma_{ac}^2 \\ \sigma_{ab}^2 & \sigma_{bb}^2 & \sigma_{bc}^2 \\ \sigma_{ac}^2 & \sigma_{bc}^2 & \sigma_{cc}^2 \end{bmatrix},
% \end{equation}
%
%Where the matrix elements of $\Sigma$ are defined by the covariance kernel
%
\begin{equation}
    \Sigma_{ij} = \mathrm{Cov}(p_i, p_j) = K(\varepsilon_i, \varepsilon_j).
\end{equation}
A common choice is the squared-exponential kernel
\begin{equation}
    K_\mathrm{se}(\varepsilon_i, \varepsilon_j) = \sigma^2 \exp\left( -\frac{(\varepsilon_i - \varepsilon_j)^2}{l^2} \right),
\end{equation}
although more complicated kernels are also used~\cite{Essick:2019ldf}.\footnote{It is worth noting that linear regression is a special case of a GP. That is, a GP can reproduce the linear model with an appropriate choice of covariance kernel: $K(\varepsilon_i, \varepsilon_j) \propto \varepsilon_i \varepsilon_j$.}

Figures~\ref{fig:toy model envelope} and~\ref{fig:toy model corner} show a GP assuming a squared-exponential kernel with parameters chosen to match the marginal distribution of the two-point prior process and $l \ll |\varepsilon_a - \varepsilon_b|$.
We also obtain
\begin{align}
    I(p_a, p_b)
        & = -\frac{1}{2}\ln \left( 1 - \frac{\sigma_{ab}^4}{\sigma_{aa}^2 \sigma_{bb}^2} \right) \nonumber \\
        & = -\frac{1}{2}\ln \left\{ 1 - \exp\left[-\frac{2(\varepsilon_a-\varepsilon_b)^2}{l^2}\right]\right\}, \label{eq:mutual information GP}
\end{align}
which vanishes as $\exp[-2(\varepsilon_a-\varepsilon_b)^2/l^2]$ in the limit $|\varepsilon_a-\varepsilon_b| \gg l$.
Furthermore, $P(p_c|p_a, p_b)$ is a normal distribution, and the generalization of the mutual information in Eq.~\eqref{mutual_info_gen} no longer diverges.
%with mean
% \begin{equation}
%     \mu_{c|a,b} = \mu_c - \frac{(\Sigma^{-1})_{ca}(p_a - \mu_a) + (\Sigma^{-1})_{cb}(p_b - \mu_b)}{(\Sigma^{-1})_{cc}}
% \end{equation}
% and variance
% \begin{equation}
%     \sigma^2_{c|a,b} = \frac{1}{(\Sigma^{-1})_{cc}}
% \end{equation}
If $l \ll |\varepsilon_b - \varepsilon_c|, |\varepsilon_a - \varepsilon_c|$, then $\Sigma_{ij} \rightarrow \sigma^2 \delta_{ij}$ and $P(p_c|p_a, p_b) \rightarrow P(p_c) \ \forall \ (p_a, p_b)$.
Therefore, if $l \ll |\varepsilon_b - \varepsilon_a|$ as well, $I(p_a, p_b, p_c) \rightarrow 0$.
We conclude, then, that the GP prior process can be made to simultaneously produce reasonable envelope plots (broad marginal distributions for all pressures) while retaining vanishingly small correlations between (reasonably separated) pressures.
This is in stark contrast to the parametrized prior processes, where this is, in general, not possible.

We demonstrate one more useful diagnostic in this toy model through the conditioned distribution
\begin{equation}
    P(p_i | p_j) = \frac{P(p_i, p_j)}{P(p_j)}
\end{equation}
which shows how our knowledge of $p_i$ depends on $p_j$.
Figure~\ref{fig:toy model envelope} shows the envelope plots for the conditioned distributions corresponding to each of our prior processes when we condition on $p_c$.
We see that a constraint at $\varepsilon_c$ is broadcast to nearby densities in all cases, but the Gaussian process fills up the prior volume from the unconditioned marginal distributions the fastest.
This is a visual manifestation of the correlations quantified by the mutual information.
Indeed
\begin{equation}
    I(a, b) = \int da P(a) \int db \, P(b|a) \ln \frac{P(b|a)}{P(b)},
\end{equation}
is just the Kullback–Leibler divergence $D_{\mathrm{KL}}\quant{P(b|a)|| P(b)}$ from the unconditioned marginal to the conditioned marginal averaged over the possible  $a \sim P(a)$.

In the case of realistic EoS inference, we have to consider even higher-dimensional spaces. A natural measure of correlations, then, is a generalization of the mutual information (sometimes called the total correlation, multivariate constraint, or multi-information~\cite{InfoTheory})
\begin{equation}
    I(x_1, \cdots, x_N) \equiv -H(x_1, \cdots, x_N) + \sum_{i=1}^N H(x_i)
\end{equation}
where $H(x) = - \int dx P(x) \ln P(x)$ is the entropy of the distribution $P(x)$.
Larger $H$ imply broader distributions.
We will consider this statistic in the context of real astrophysical constraints on the EoS in Sec.~\ref{interdensity-correlations}.
In general, one can make $I$ small but still allow for very little model freedom (small $H$).
We therefore seek prior processes with both large $H(x_1,\cdots, x_N)$ and small $I(x_1, \cdots, x_N)$.
This is sometimes captured in the variation of information, defined as $H - I$, but we find it more useful to consider $H$ and $I$ separately.

%%%%%%%%%%%%%%%%%%%%%%%%%%%%%%%%%%%%%%%
\section{Impact of interdensity correlations: idealized measurement}
%\section{The effect of interdensity correlations on EoS Inference}
\label{interdensity-correlations}
%%%%%%%%%%%%%%%%%%%%%%%%%%%%%%%%%%%%%%%

\begin{figure*}
    \centering
    \includegraphics[width=.49\textwidth]{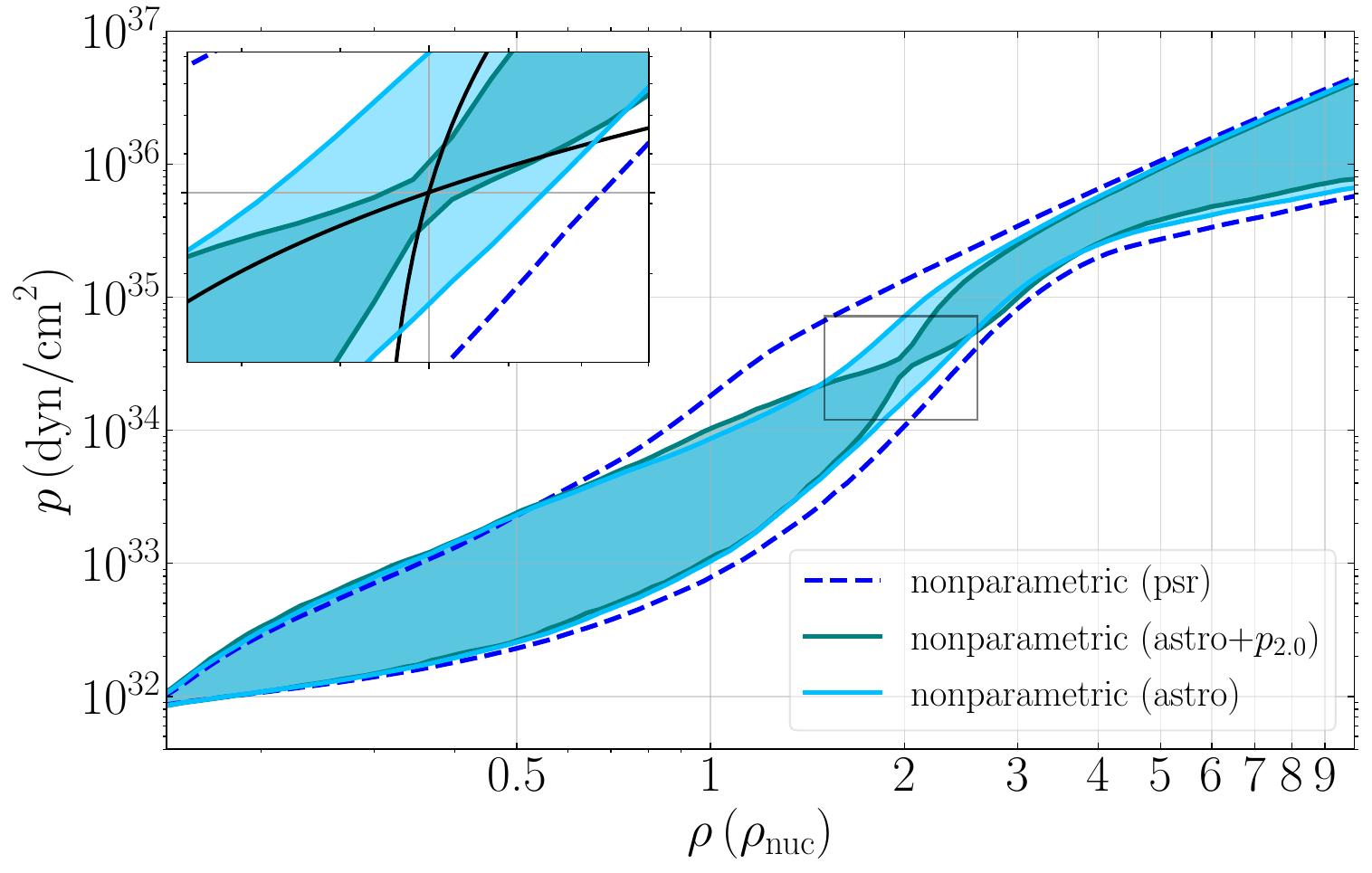}
    \includegraphics[width=.49\textwidth]{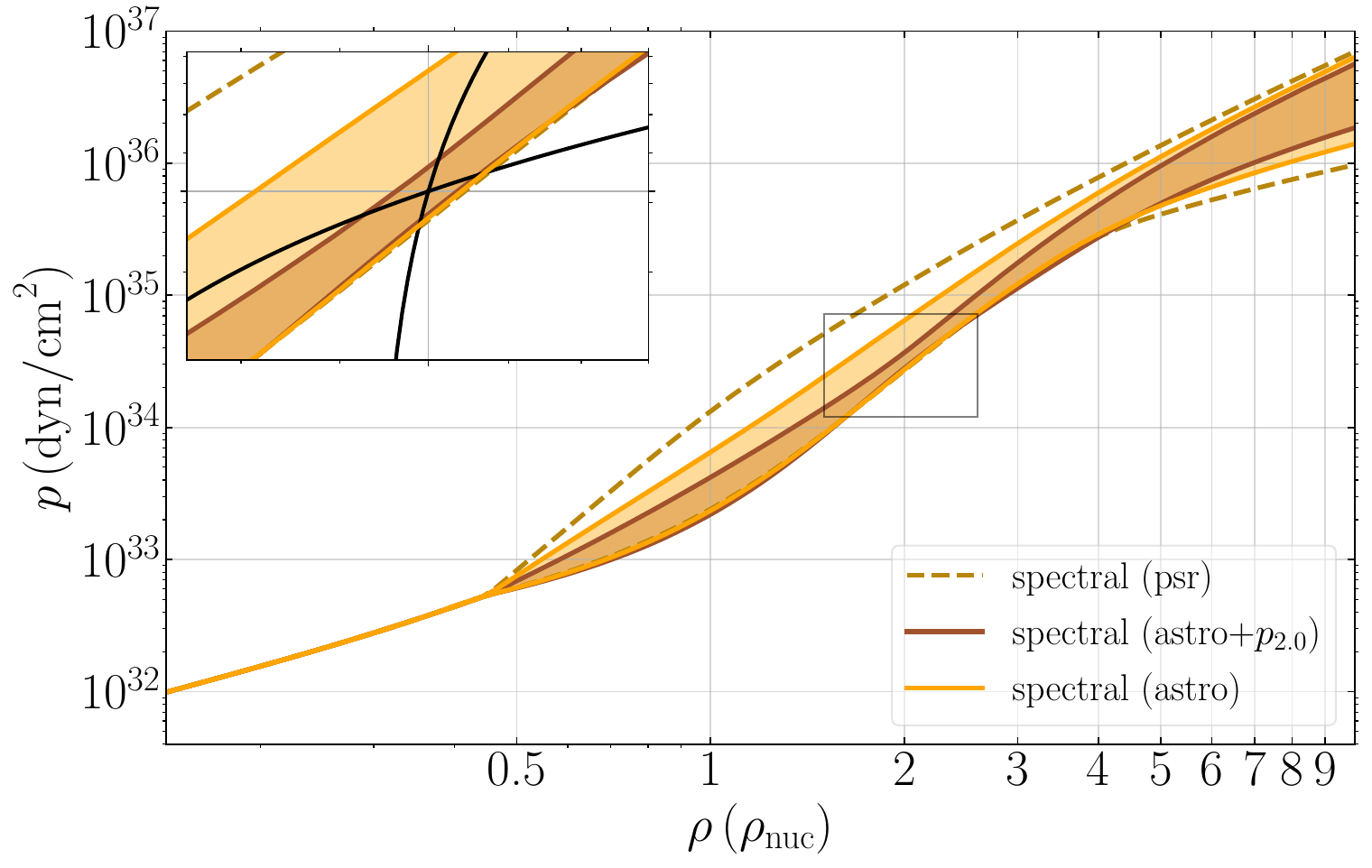}
    \includegraphics[width=.49\textwidth]{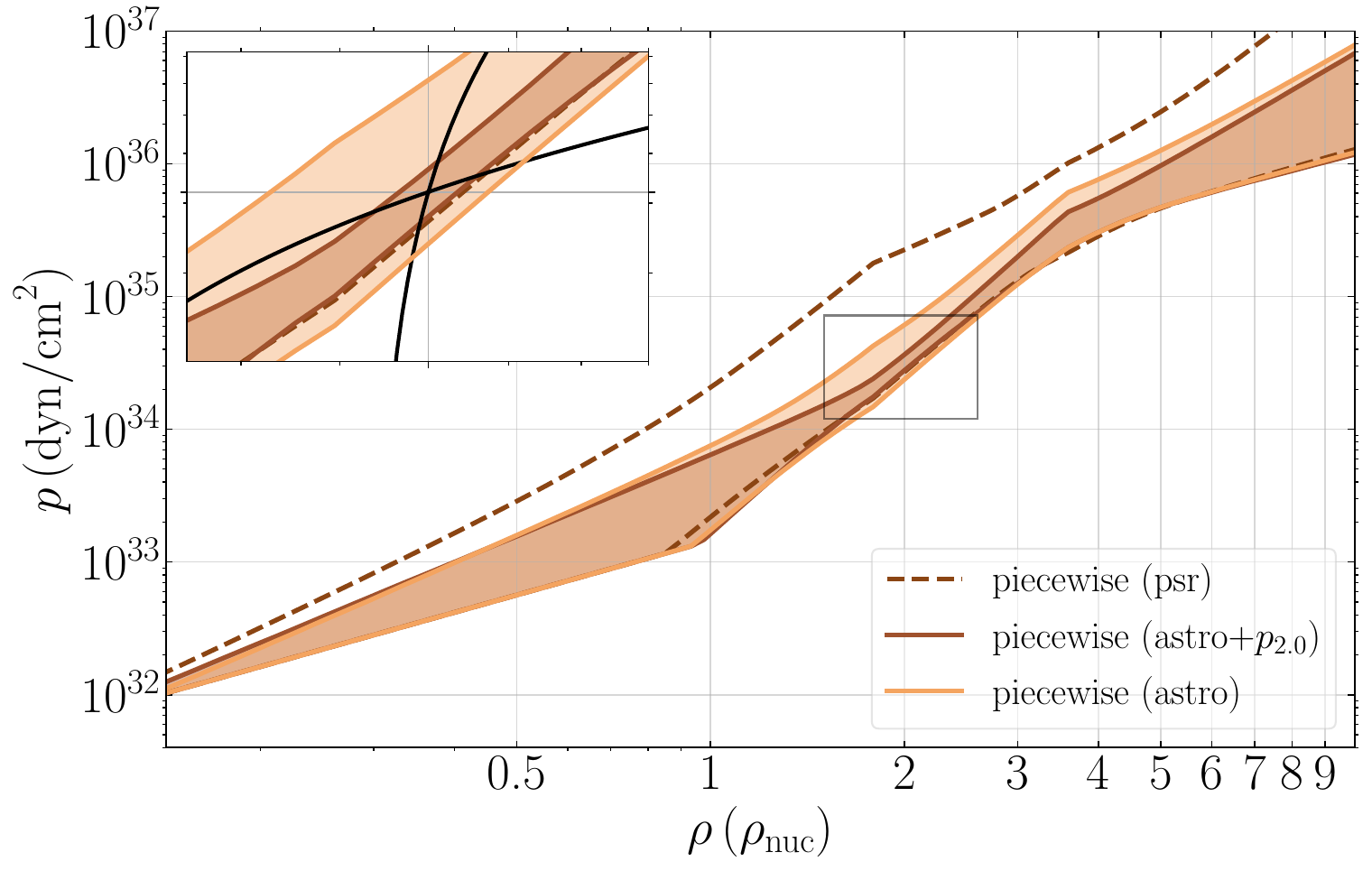}
    \includegraphics[width=.49\textwidth]{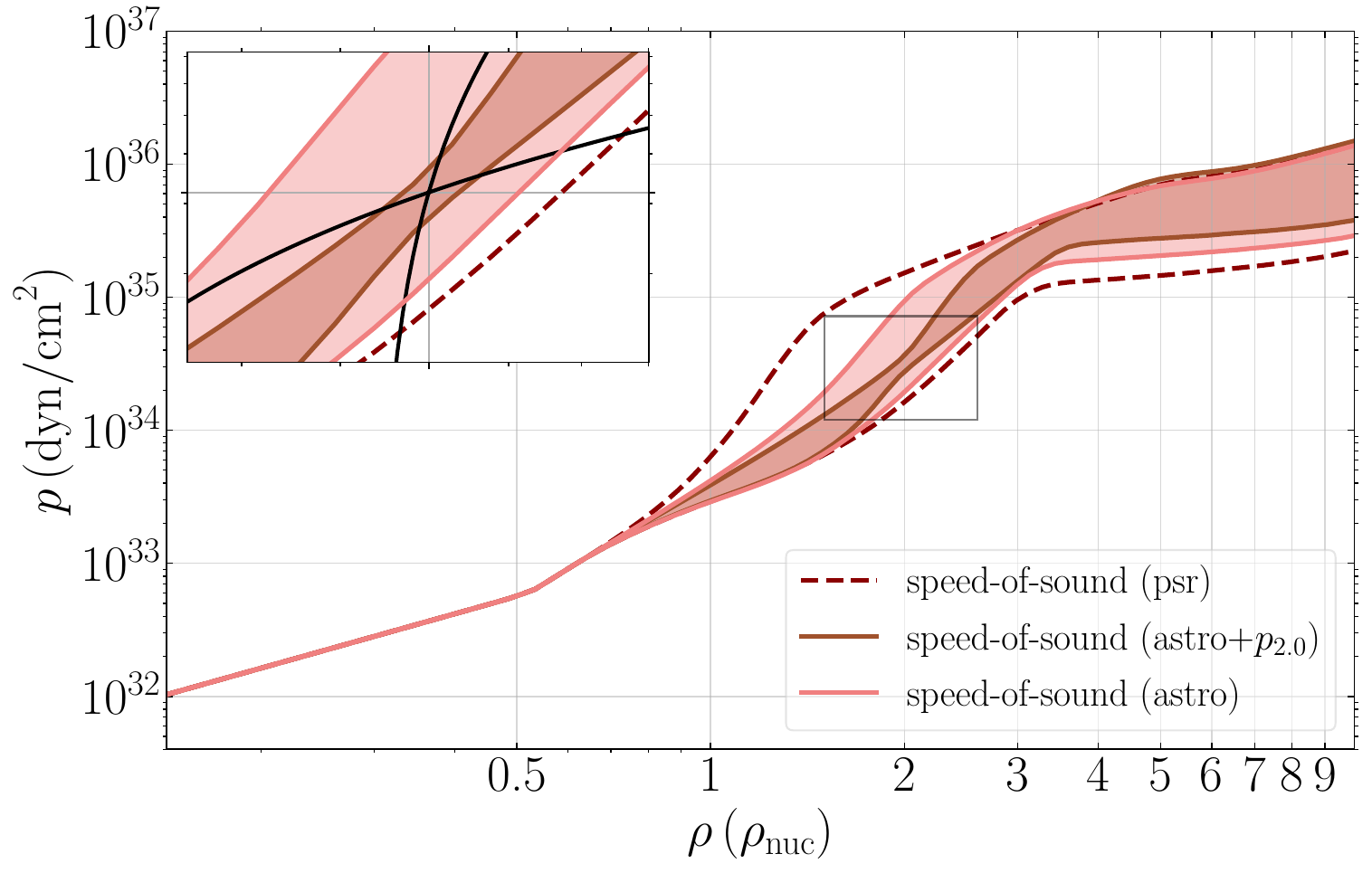}
    \caption{
        Similar to Fig.~\ref{fig:process_prior_plot} but with a mock constraint injected directly into $p(\rho=2\rho_{\nuc})$ for each EoS prior process.
        The posterior after the simulated constraint is included (``astro+mock") is overplotted on the posterior with all current data (``astro"), and in a darker color.
        The constraint at a single density affects the parametric posteriors over a much wider range of density scales than the nonparametric one.   
        The inset focuses around $\rho=2\rho_{\nuc}$.
        The two black straight lines provide an estimate of the constraints imposed by causality ($c_s^2<1$) and thermodynamic stability $(c_s^2>0.1)$ around $\rho=2\rho_{\nuc}$, subject to the heavy pulsar measurements.
        The nonparametric posterior quickly ``fills" more of the physically available region after satisfying the mock constraint, while the parametric posteriors do not.  \\
    }
    \label{fig:mock_pressure_injections_comparison}
\end{figure*}

Our toy model illustrates the potential impact of implicit correlations within EoS models on the results of EoS inference. In order to quantify the sensitivity of the parametric and nonparametric models to such model-dependent correlations between density scales, we now consider simulated NS observations.
The macroscopic NS properties that astronomical observations target (masses, radii, and tides) are determined by a range of NS densities, it is therefore not straightforward to disentangle the effect of the data and the model dependence in the constraint that a single astronomical observation imposes on the EoS.
Consequently, we begin with the same setup as Sec.~\ref{sec:toy model}: an idealized direct measurement of $p(\rho)$ at a single density, while keeping in mind that a realistic astronomical measurement would correspond to a combination of many such constraints correlated across many densities.\footnote{An example of how one may obtain direct constraints on the pressure from nuclear experiments is demonstrated in Refs.~\cite{Essick:2021ezp,Essick:2021kjb}.}

We consider a tight Gaussian constraint at $p_{2.0}$ with mean of $3.20 \times 10^{34}\, \mathrm{dyn}/\mathrm{cm}^2$ based on a candidate EoS drawn from our GP prior that is consistent with all current parametric posteriors near $2\rho_{\nuc}$.
We arbitrarily choose the standard deviation, $2.61\times 10^{33}\,\mathrm{dyn}/\mathrm{cm}^2$ ($\sim8\%$ relative uncertainty).
We then plot the corresponding envelope for $p(\rho)$ with this mock constraint and all other real astronomical data for each model in Fig.~\ref{fig:mock_pressure_injections_comparison}.
In the nonparametric case, imposing this constraint pinches the $p$-$\rho$ envelope around $2\rho_{\nuc}$, but the uncertainty in the $p(\rho)$ curve is unaffected beyond $\approx \pm 0.5 \rhonuc$.
All the parametric models, though, change across several $\rhonuc$, indicating that the EoS at many scales is informed significantly by the EoS near $2\rhonuc$.  

In each panel, the inset zooms in around the $\rho=2\rho_{\nuc}$ region; to guide the eye the two black lines provide a rough estimate of the maximally causal ($c_s^2=1$) and minimally stable $(c_s^2=0.1)$ EoS that can support the heavy pulsar observations (see Fig. 2 of Ref.~\cite{Landry:2020vaw}) around $\rho=2\rho_{\nuc}$.
The two lines were obtained by combining $dp/d\varepsilon=c_s^2$ and the first law of thermodynamics with the approximation that $\varepsilon_{2.0}=c^2\rho_{2.0}$.
The nonparametric prior process contains EoS draws that approach this limiting behavior near the constraint.
Comparing the $4$ panels, the nonparametric model fills more of the physically available space.
The parametric models, on the other hand, are clearly subject to additional correlations between pressures besides causality and stability.

To quantify these correlations, we follow Sec.~\ref{sec:toy model} and compute the total correlation between the pressures at several reference densities.
Table~\ref{tab:total correlation} shows the total correlation ($I$) and joint entropy ($H$) between $\ln\,p_{1.0}$, $\ln\,p_{1.5}$, $\ln\,p_{2.0}$, $\ln\,p_{3.0}$, and $\ln\,p_{4.0}$ induced by the posterior process conditioned on the astrophysical data as well as the astrophysical data and the mock constraint on $p_{2.0}$.\footnote{We estimate the entropies via Monte Carlo sums over kernel density estimates (KDEs) of the associated distributions. As such, the actual correlations may be smoothed by the KDE, which may act as upper limits on the estimates of the mutual information in some cases.}
We consider these pressures as the central density of $\Mmax$ stars may be as low as $4\rhonuc$~\cite{Legred:2021}, and therefore we focus on pressures that are confidently relevant for NSs.
Although the precise values of $I$ and $H$ can be difficult to interpret, we notice some trends.

\begin{table*}
 \centering
 \caption{
     Total correlation ($I$) and entropy ($H$) of the joint distributions over $\ln\,p_{1.0}$, $\ln\,p_{1.5}$, $\ln\,p_{2.0}$, $\ln\,p_{3.0}$, and $\ln\,p_{4.0}$ induced by several processes as well as the entropy of the marginal distribution over only $\ln\, p_{2.0}$ ($H(\ln\,p_{2.0})$).
     The nonparametric processes consistently have smaller $I$ and (much) larger $H$ than any parametric process, implying much more model freedom.
     This is the case even though the entropy of the marginal distributions for $\ln\,p_{2.0}$ can be comparable.
 }
 {\renewcommand{\arraystretch}{1.2}
 \begin{tabular}{@{\extracolsep{0.25cm}} c ccc | ccc | ccc}
    \hline
    \hline
      & \multicolumn{3}{c}{$I$} & \multicolumn{3}{c}{$H$} & \multicolumn{3}{c}{$H(\ln p_{2.0})$} \\
      \cline{2-10}
      & PSR & Astro & Astro+$p_{2.0}$ & PSR & Astro & Astro+$p_{2.0}$ & PSR & Astro & Astro+$p_{2.0}$ \\
    \hline
    Nonparametric
      & 3.7 & 3.1 & 2.9 & 0.7 & -1.0 & -2.5 & 1.0 & 0.5 & -1.1 \\
    Spectral
      & 6.6 & 5.5 & 4.7 & -4.2 & -5.5 & -7.6 & 0.5 & 0.0 & -1.1 \\
    Polytrope
      & 5.7 & 4.6 & 3.8 & -1.6 & -3.6 & -5.7 & 0.9 & 0.2 & -1.1 \\
    Speed of sound
      & 5.0 & 4.7 & 4.3 & -2.6 & -4.3 & -7.1 & 1.0 & 0.6 & -1.1 \\
    \hline
 \end{tabular}
 }
 \label{tab:total correlation}
\end{table*}

Overall, the nonparametric process consistently has the largest joint entropy and smallest total correlation, as desired.
The parametric processes all have approximately equal $I$, which are larger than the nonparametric process by $\gtrsim 1$ nat.
Additionally, the change in $I$ when we additionally condition on a mock constraint on $p_{2.0}$ is much smaller for the nonparametric than for the parametric processes.
This can be interpreted as the constraint on $p_{2.0}$ removing some correlations from the parametric processes by approximately fixing the value of $p_{2.0}$.

What is more, the parametric processes have much smaller joint entropies than the nonparametric process in all cases.
This is a manifestation of the reduced model freedom in the parametric processes as the nonparametric process explores more combinations of pressures than any parametric process.
Although not exact, the exponential of the difference in entropies is an estimate of the ratio of the effective number of pressure combinations supported in each distribution: the nonparametric contains between 10 and 100 times as many possible pressure combinations as the parametric processes.

Additionally, the constraint on $p_{2.0}$ removes more entropy from each parametric processes than from the nonparametric process.
While we expect the joint entropy to be smaller in all cases after measuring $p_{2.0}$ precisely, the additional entropy lost in the parametric processes is associated with the correlations between pressures.
That is, knowledge of $p_{2.0}$ decreases our uncertainty in other pressures within the parametric processes, something that does not happen as strongly in the nonparametric process.
This is apparent in Fig.~\ref{fig:mock_pressure_injections_comparison} as well. 

As a final note, Table~\ref{tab:total correlation} also reports the entropy of the marginal distributions over $\ln p_{2.0}$.
We see smaller differences between these one-dimensional~(1D) distributions, reinforcing the conclusion that the differences between the nonparametric and parametric processes arise mainly from correlations between multiple pressures.

%%%%%%%%%%%%%%%%%%%%%%%%%%%%%%%%%%%%%%%%%%%%%%%%%
\section{Impact of interdensity correlations: Mock astrophysical observations}
\label{mock-astro}
%%%%%%%%%%%%%%%%%%%%%%%%%%%%%%%%%%%%%%%%%%%%%%%%%

Different astronomical probes provide information about different density scales, and therefore interdensity correlations are likely to matter even more for realistic EoS inference than in the idealized case considered above. %will matter when we compare their results.
Implicit correlations in EoS models could artificially give the appearance of tension between observations of NSs or nuclear matter made via different channels.
This has already been shown to be relevant in comparisons of nuclear experiments with astrophysical observations~\cite{Essick:2021ezp, Essick:2021kjb}. 
To investigate this possibility, we now repeat the previous sections' analysis for a simulated set of astronomical observations.

We choose a candidate EoS with $\Mmax=2.54\, \Msolar$ and $\Rtyp=12.0\, \mathrm{km}$.  
This EoS is relatively soft at low densities and stiff at high densities, but is consistent with the $90\%$ 1D marginal pressure constraints for all of our models at all densities, except the speed-of-sound parametrization above $4 \rhonuc$.  The combination of macroscopic parameters lies outside the spectral 90\% credible region in Fig. \ref{fig:spectral_Mmax-Rtyp}, motivating its use in studying how tension appears in an analysis when such a mismatch arises.\footnote {Due to the broad prior of the nonparametric model, finding a physically valid EoS with no support in the nonparametric macroscopic or microscopic priors is much more challenging.}
We simulate three measurements of pulsar masses and radii with comparable uncertainty to the recent measurement for J0740+6620~\cite{Miller:2021qha, Riley:2021pdl}.
This observation incorporated radio data to constrain the pulsar mass \cite{Fonseca:2021wxt} and constrained the radius with x-ray data.  We also simulate 20 GW detections of binary NS mergers at A+ detector sensitivity~\cite{Aasi:2020wya}.
Note, however, that we do not impose prior knowledge of the NS nature of the components in our inference.
The simulated pulsars are drawn from a uniform-in-central-density distribution, while the simulated binary NSs come from a uniform-in-mass distribution, under the condition that the NS masses lie below $\Mmax$.

\begin{figure}
    \centering
    \includegraphics[width=.49\textwidth]{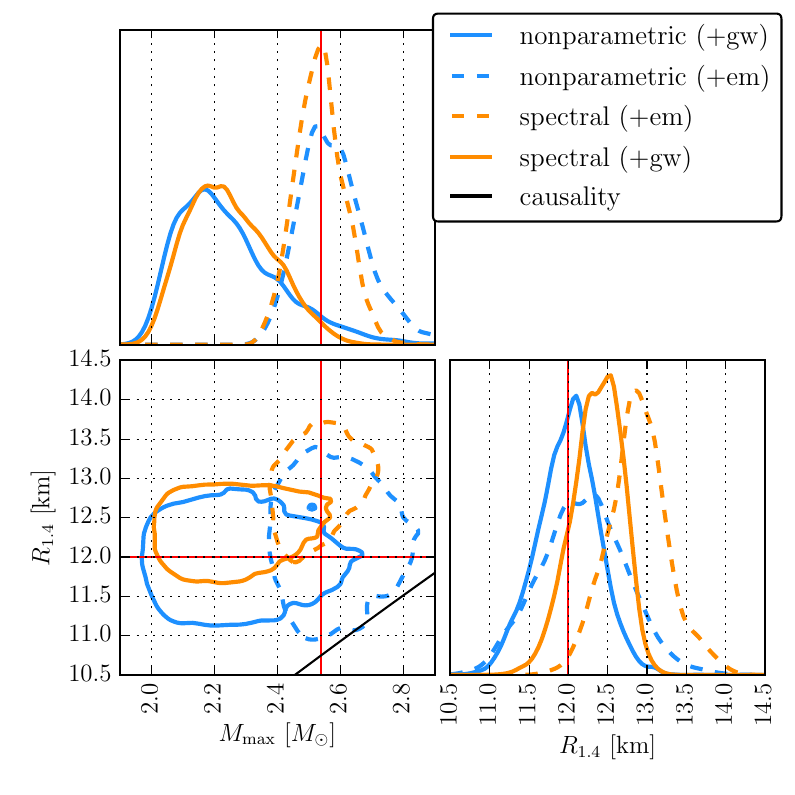}
    \caption{
        Inferred posterior for $\Rtyp$ and $\Mmax$ using the nonparametric model and the spectral model and mock x-ray-radio (blue and orange dashed line, respectively) and GW (blue and orange solid line, respectively) observations.  All  posteriors also include all current astrophysical data.
        The vertical and horizontal red lines show the injected value of $\Rtyp$ and $\Mmax$. The $\Mmax$ posterior is only weakly informed by the GW data as they typically cannot lead to a definitive identification of a $>2M_{\odot}$ object as a NS, and it is thus similar to that of Fig.~\ref{fig:spectral_Mmax-Rtyp}.
    }
    \label{fig:rtyp-tension}
\end{figure}

We analyze each dataset separately, folding the simulated measurements of each type onto all current astrophysical data.  We plot the inferred posteriors in Fig.~\ref{fig:rtyp-tension}.
We find that the nonparametric posterior for $\Rtyp$ is centered on the correct value ($12\,\mathrm{km}$) with either x-ray or GW data.
In the spectral case, though, while the GW measurements are consistent with the correct value, the x-ray posterior is in tension at $90\%$ credibility.
Moreover, the GW and x-ray posteriors are less consistent with each other, an observation that could lead to the erroneous conclusion of tension between different EoS probes.

We can understand this as follows.
We expect GW measurements of high-mass NSs ($\geq 1.7 \Msolar$) to be less informative than lower-mass NSs, as the absolute impact of tidal parameters on the signal is weaker for more compact stars.
In general, high-mass NSs  will most likely be indistinguishable from black holes until the advent of next-generation detectors~\cite{Chen:2020fzm, Legred:2021}.
As such, high-mass systems offer little information for either $\Mmax$ or $\Rtyp$, and thus the GW data primarily probe only the low-mass/low-density part of the EoS.
Indeed, Fig.~\ref{fig:rtyp-tension} shows that each mock-GW $\Mmax$ posterior is similar to the respective posterior of Fig.~\ref{fig:spectral_Mmax-Rtyp},  indicating that additional GW observations inform $\Mmax$ only weakly.

On the other hand, x-ray measurements have already proven capable of bounding the radius of high-mass NSs \cite{Miller:2021qha, Riley:2021pdl}.
Additionally, x-ray detection of pulsations in a compact object proves it is a NS, and thus its mass offers information about $\Mmax$.
Depending on the mass distribution of observed events, x-ray probes could thus probe the EoS at both low and high densities.
Our mock x-ray dataset contains one such  NS with mass $2.50\,M_{\odot}$.
Figure~\ref{fig:rtyp-tension}, then, shows that when we use a parametric model to fit all the x-ray data, biases can arise as no EoS in the prior process can simultaneously reproduce the correct values for both $\Mmax$ and $\Rtyp$.
The bias is smaller in the GW-based results as the data there probe a narrower density range, resulting in the appearance of mild tension between the two datasets. 
By extension, a newly observed GW signal for a 1.4$M_{\odot}$ NS  would be in tension with the x-ray-based results, despite no real astrophysical inconsistency.
Recent concerns of tensions between PREX-II~\cite{Reed:2021nqk} and astrophysical predictions may be influenced by a similar mechanism, as noted in Refs.~\cite{Essick:2021ezp, Essick:2021kjb}.

%%%%%%%%%%%%%%%%%%%%%%%%%%%%%%%%%%%%%%%
\section{Impact of parametric prior choices}
\label{parametric-prior-choices}

These investigations show that the parametric EoS prior processes include model-dependent interdensity correlations that influence the resulting inference.
Such prior processes are constructed based on two ingredients: (i) a functional form for $p(\rho)$ (or an equivalent quantity) and (ii) a prior for the parameters of the function.
The former may be carefully engineered, while the latter can be changed more easily.
 As in Sec.~\ref{sec:toy model}, it is therefore reasonable to wonder if we can change the nature of the interdensity correlations by a trivial change in the parameter prior, or whether the correlations are inherent to the functional form. Below we argue for the latter, as also demonstrated in Sec.~\ref{sec:toy model}.

First, adding parameters does not necessarily always increase model freedom.
Adding a parameter to any model we have shown so far will require choosing a distribution for that parameter, and a reasonable range will strongly depend on the functional form.
For the piecewise-polytrope model, this process is somewhat easier, as additional adiabatic indices for new segments have clear physical meaning. Therefore reasonable ranges can be chosen.
For the spectral model, with the addition of a new spectral component, there is no obvious mapping of parameter values to physics, and so tuning parameter ranges is much harder. 

Second, of the existing parameters in the models, we typically find that only a few are meaningfully constrained.
For example, the speed-of-sound model has only a single Gaussian bump, and thus current astrophysical data tightly constrain this bump to be at densities low enough to produce pulsars consistent with, e.g., Refs.~\cite{Miller:2019nzo, Riley:2019yda, Landry:2020vaw}.
This severely limits the flexibility of the model, as the logistic term is, by itself, not strong enough to support realistic NSs. In practice $a_1$ and $a_2$ are overconstrained in this model and $a_4$, and $a_5$ are underconstrained.
As a result, the speed-of-sound model is the least flexible (and leads to the most stringent constraints) even though it has the most parameters.
We find similar behavior in the spectral and piecewise-polytrope models, suggesting that the effective number of parameters in the models is  fewer than what is nominally stated.

\begin{figure*}
    \centering
    \includegraphics[width=0.9\textwidth]{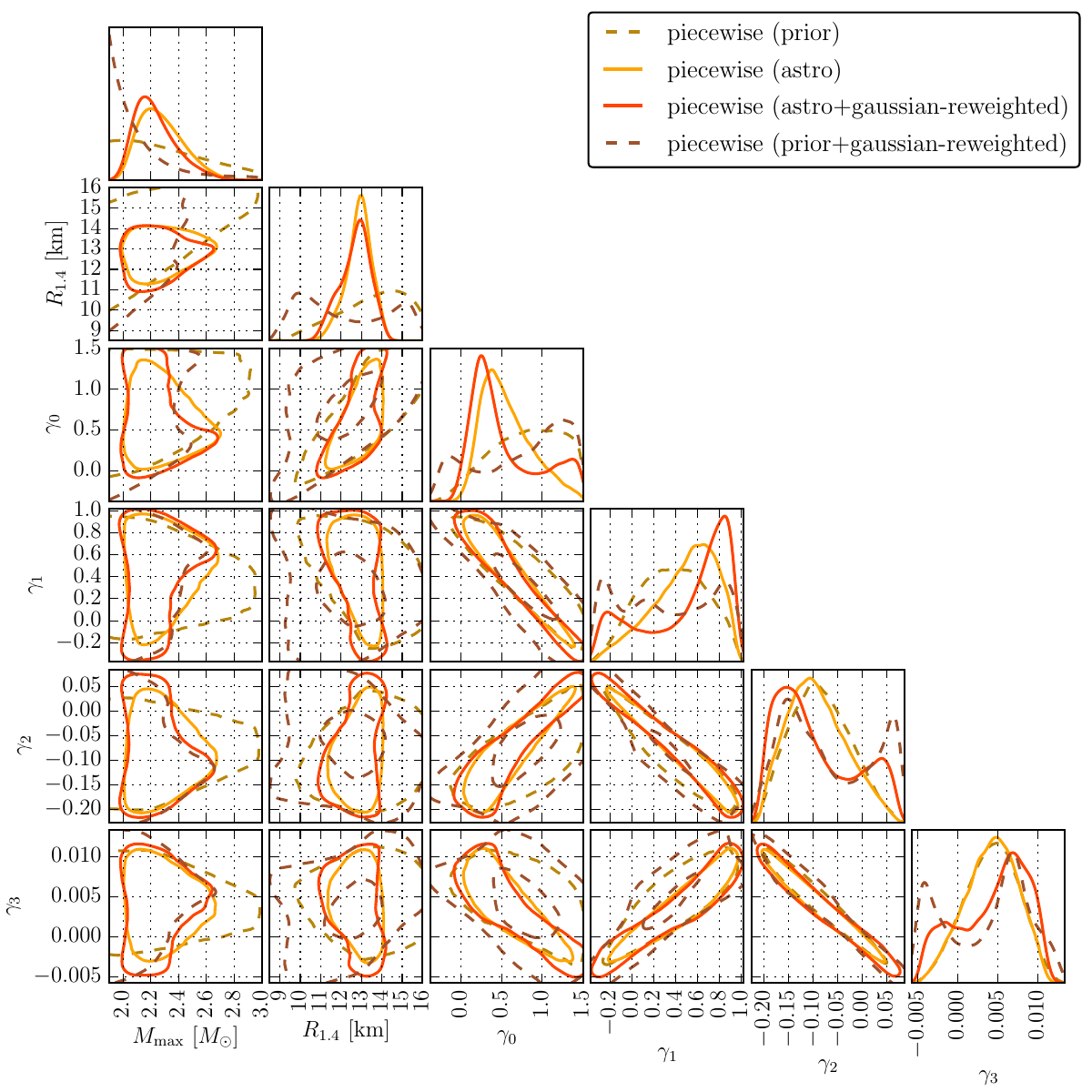}
    \caption{
        Marginal one- and two-dimensional prior and posterior distributions for the parameters of the spectral model, $\gamma_i$, as well as the maximum NS mass, $\Mmax$, and the radius, $\Rtyp$.
        We show the default prior as well as reweighted results that upweight more extreme values of $\gamma_i$.
        In both cases, $\Mmax$ and $\Rtyp$ have very similar posteriors, showing that the reweighting does not efficiently extend the coverage of the prior toward the causality threshold in the $\Mmax$-$\Rtyp$ plane.
        Additionally, extending the prior ranges for $\gamma_i$ is unlikely to change the results as the posteriors are not limited by the prior ranges. 
    }
    \label{fig:spectral_reweighted_params_corner}
\end{figure*}

Third, due to the fine-tuning of the parametric models, attempting to redefine priors on parameters is generally not an efficient way to expand model freedom.
As an example, we consider the spectral EoS where we find that EoS candidates have \textit{a priori} strong correlations between parameters in order to satisfy causality and stability.
These correlations were noted in Ref.~\cite{Wysocki:2020myz} and are also shown in Fig.~\ref{fig:spectral_reweighted_params_corner}. We find that the $\gamma_i$ are alternately strongly correlated or anticorrelated with each other.
It is possible that other distributions, in particular distributions that upweight EoS further from the line of strongest correlation, reduce the strength of interdensity correlations.  

We test this by upweighting more extreme $\gamma_i$ values.
The reweighting procedure does indeed change the posterior distribution of $\gamma_i$ parameters, although the inferred distribution in $\Mmax$-$\Rtyp$ is effectively unchanged.
Moreover, the $\Mmax$-$\Rtyp$ prior does not significantly extend into the previously excluded region closer to the stability threshold.
We find similar results with a different reweighting of $\gamma_i$ that instead favors for central values.
Figure~\ref{fig:spectral_reweighted_params_corner} also shows that extending the prior range on $\gamma_i$ will not extend the reach of the spectral model as the parameter posteriors are not significantly affected by the prior cutoffs.
We reach similar conclusions with the piecewise polytrope.

Overall, while it was possible to remove correlations between only a subset of pressures within our toy models in Sec.~\ref{sec:toy model}, it is generally difficult to do even that with real parametrized EoS models.

 %%%%%%%%%%%%%%%%%%%%%%%%%%%%%%%%%%%%%
\section{Conclusions}
\label{discussion}
 %%%%%%%%%%%%%%%%%%%%%%%%%%%%%%%%%%%%%

All models of the dense-matter EoS should contain some correlations between density scales due to causality and thermodynamic stability requirements.
However, in this study we show that phenomenological parametric models such as the spectral, piecewise-polytrope, and speed-of-sound models impose even stronger correlations \textit{a priori}.
As a result, NS properties are constrained more tightly in parametric models than in nonparametric ones in ways that are not supported by the data.
Regardless of whether these tighter constraints end up being compatible with the true EoS, their emergence is attributable to what are effectively model-dependent prior assumptions dictated by the phenomenological nature of the parametrizations.
Viewed in this way, they deserve the same scrutiny as other prior choices imposed by the analyst.
%Regardless of whether these tighter constraints end up being compatible with the true EoS, their emergence through model dependence is completely arbitrary, given that these parametrizations are phenomenological. That is, they are prior choices imposed by the analyst.

The concerns about implicit correlations are alleviated by GP-based nonparametric models that enjoy extensive model freedom, restricted only by causality and thermodynamic stability.
They allow us to generate, with no additional modeling effort, candidate EoS with complex phenomenology that could be associated with, e.g., a transition to quark matter in the cores of NSs.
For example, Refs.~\cite{Tan:2020ics,Tan:2021ahl} study EoS with complex speed-of-sound phenomenology, while Refs.~\cite{Han:2018mtj,Chatziioannou:2019yko,Han:2020adu,Drischler:2020fvz,Li:2021crp,Drischler:2021bup} consider strong first-order phase transitions that result in a discontinuity in the speed of sound and multiple stable branches.
The GP prior process is able to recreate such behaviors generically.

The parametric models are relatively easier to implement.
However this might come at the cost of fine-tuning which makes it harder to sample from the prior as many draws are unphysical.
Extension to more complex phenomenology, such as phase transitions, is less straightforward and might need tailored parametric models~\cite{Alford:2013aca,Alford:2015gna,Han:2019bub} unless the parametrization supports such behavior inherently.
The piecewise polytrope specifically, as implemented here, can lead to priors and posteriors with ``kinks"~\cite{Carney:2018sdv, Lackey:2014fwa}, while Refs.~\cite{Raaijmakers:2018bln,Riley:2018ekf} discuss its behavior in cases where the observed NSs do not reach high enough densities to probe all polytropic segments. 
More complicated parametric models exist (see Appendix~\ref{parametric-eos-appendix}), but as we argue in Sec.~\ref{parametric-prior-choices}, improving parametric models by adding parameters or extending the priors ranges is not always straightforward or efficient. However, extreme extensions to these models (for example a $\mathcal{O}(1000)$ parameter extension to the piecewise polytrope) could exhibit behavior that is closer to the nonparametric results than the few-parameter models they generalize.

In conclusion, commonly used parametric models of the EoS are hampered by built-in and often opaque correlations between density scales. 
These correlations already affect inferences based on these models, and these effects will only become more severe with additional astrophysical data.
The impact of the EoS model on inference acts as an additional systematic error that must be addressed to achieve highly informative EoS constraints~\cite{Dudi:2018jzn,Gamba:2020wgg,Chatziioannou:2021tdi,Pratten:2021pro,Kunert:2021hgm,Essick:2021ezv}.
Our work shows that the nonparametric GP-based model addresses this EoS model systematic and restores model freedom by forgoing the use of specific functional forms for the EoS itself and instead parametrizing a wide range of possible correlations directly.

%%%%%%%%%%%%%%%%%%%%%%%%%%%%%%%%%%%%%
\acknowledgements

We thank Les Wade for useful discussions on the implementation of the spectal model in LAL\textsc{suite} .
R.E. thanks the Canadian Institute for Advanced Research (CIFAR) for support.
Research at Perimeter Institute is supported in part by the Government of Canada through the Department of Innovation, Science and Economic Development Canada and by the Province of Ontario through the Ministry of Colleges and Universities. P.L. is supported by the Natural Sciences and Engineering Research
Council of Canada (NSERC).

This research has made use of data, software and/or web tools obtained from the Gravitational Wave Open Science Center (https://www.gw-openscience.org), a service of LIGO Laboratory, the LIGO Scientific Collaboration and the Virgo Collaboration.
Virgo is funded by the French Centre National de Recherche Scientifique (CNRS), the Italian Istituto Nazionale della Fisica Nucleare (INFN) and the Dutch Nikhef, with contributions by Polish and Hungarian institutes.
This material is based upon work supported by NSF's LIGO Laboratory which is a major facility fully funded by the National Science Foundation.
The authors are grateful for computational resources provided by the LIGO Laboratory and supported by National Science Foundation Grants No. PHY-0757058 and No. PHY-0823459.
%

%%%%%%%%%%%%%%%%%%%%%%%%%%%%%%%%%%%%%%%%%%%%%%%%%%%
\appendix

%%%%%%%%%%%------------------------------------------------

\section{Description of the nonparametric EoS model}
\label{np-appendix}

\begin{figure}
    \centering
    \includegraphics[width=.49\textwidth]{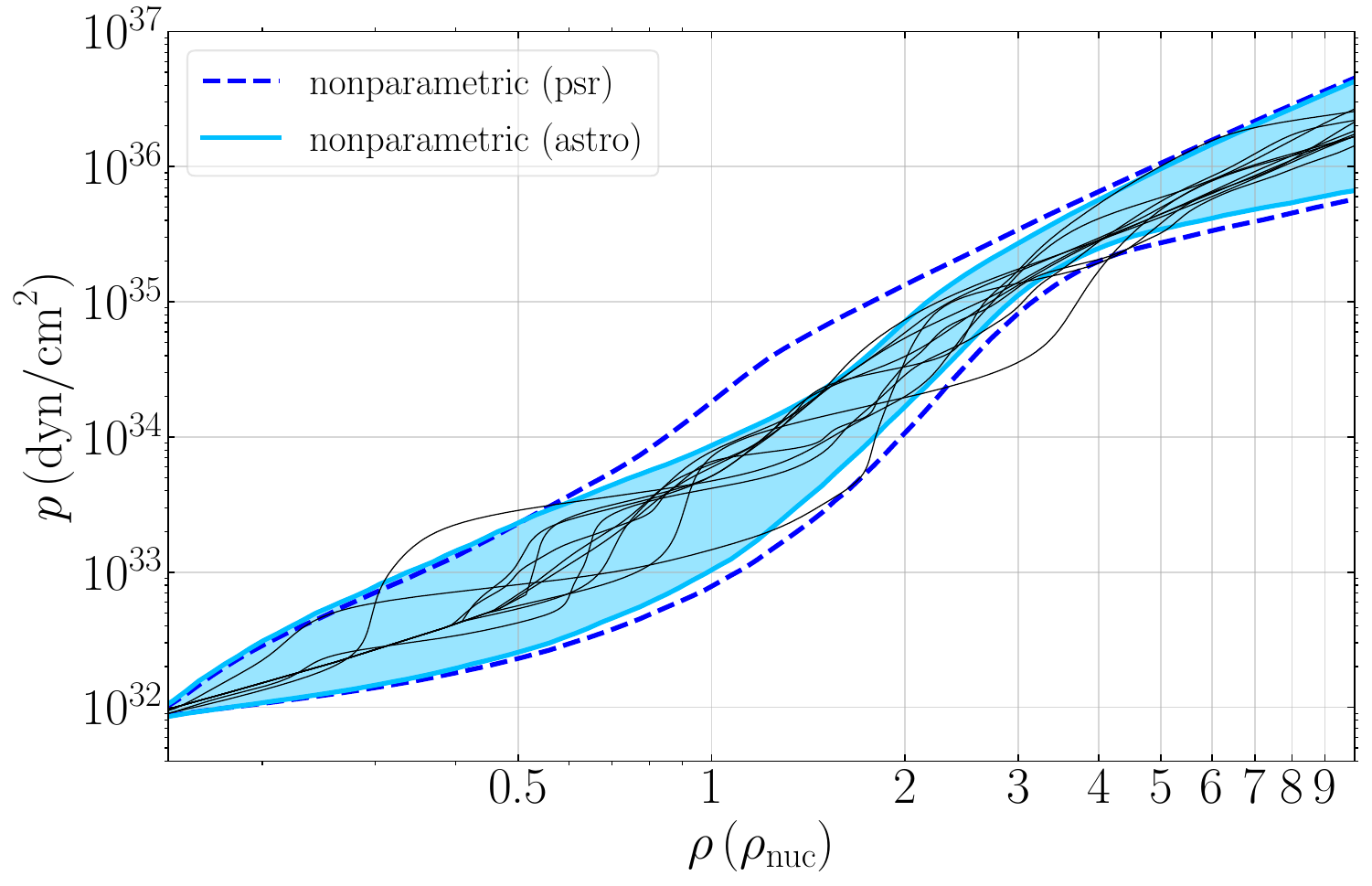}
    \includegraphics[width=.49\textwidth]{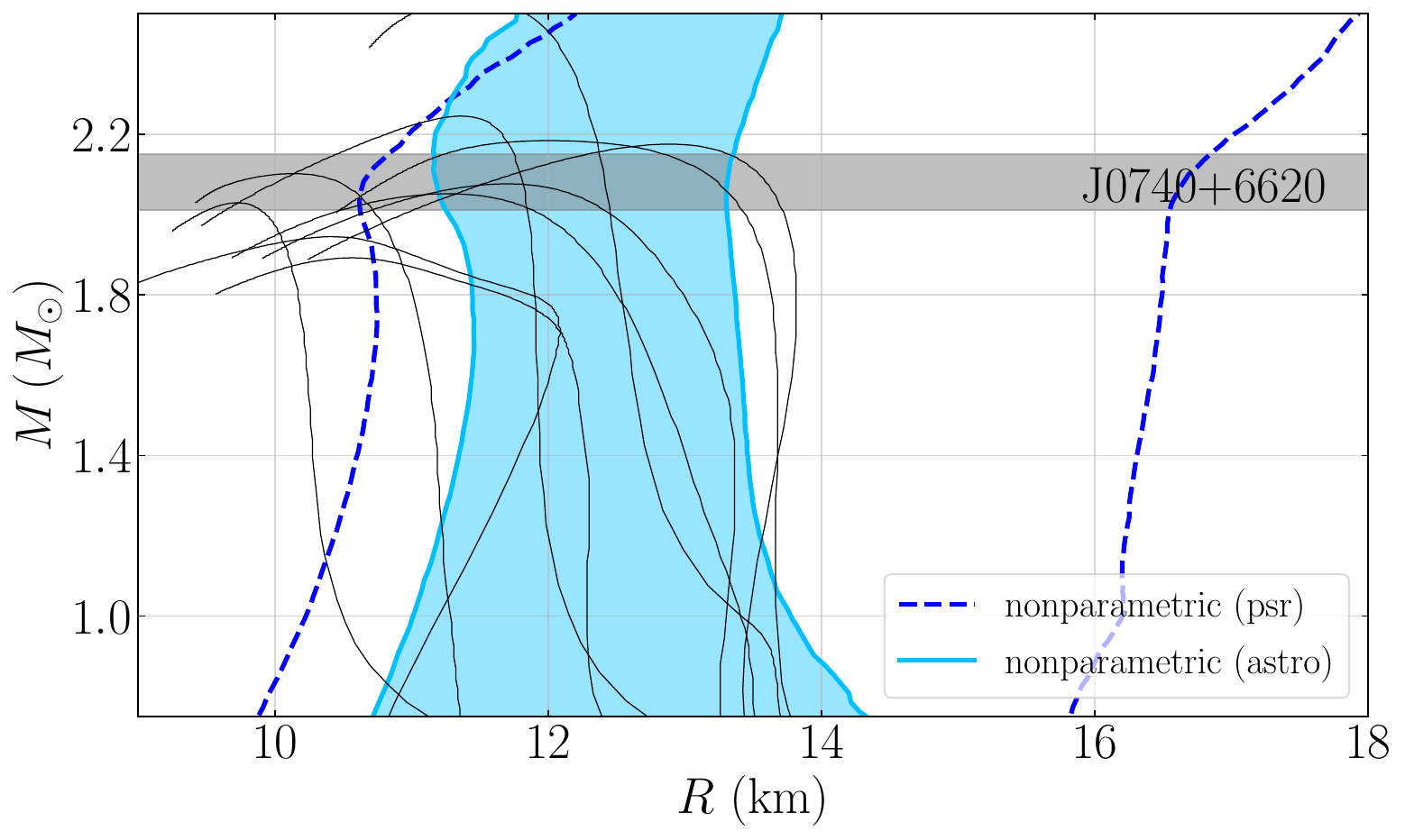}
    \includegraphics[width=.49\textwidth]{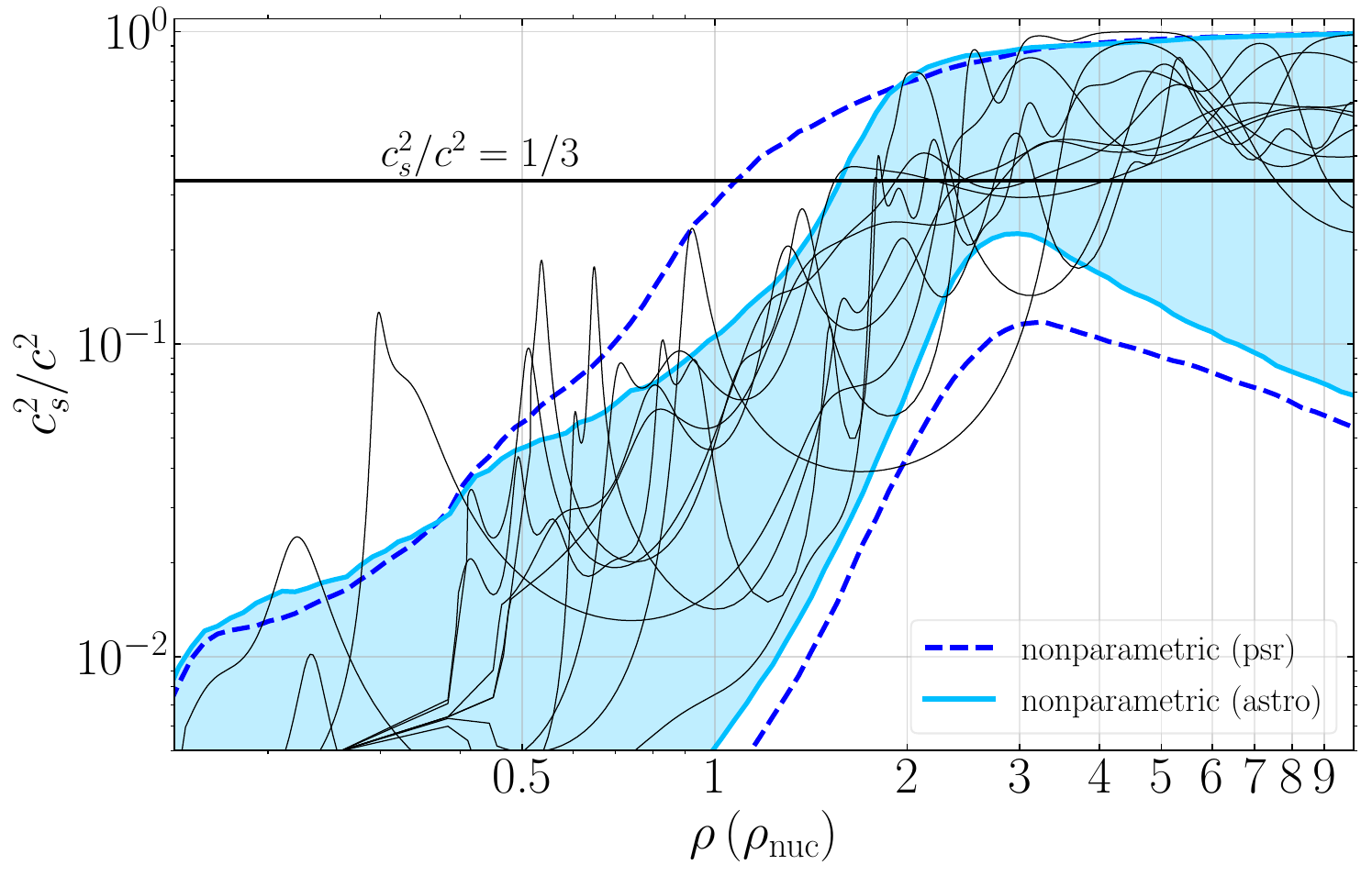}
    \caption{
        Example EoS draws from the nonparametric prior plotted in terms of the pressure $p$ vs the density $\rho$ (top), the mass $M$ vs the radius $R$ (middle), and the speed of sound $c_s$ vs the density $\rho$ (bottom).
        We only draw EoS with non-negligible contribution to the posterior.
        For reference, we also plot the 90\% symmetric credible intervals for the posterior using only heavy pulsar observations and all astrophysical data.
    }
    \label{fig:GP_example_draws}
\end{figure}

Our GP is tailored to incorporate a variety of possible correlation lengths, established by the form of the kernel function.  
Each GP draw is a realization of a multivariate Gaussian distribution, which is loosely conditioned on nuclear models.
The GP from which EoS candidates are sampled has a covariance which is governed by a kernel function through the parameters $\sigma$ and $\ell$ that control the strength and length of the correlations respectively~\cite{Essick:2019ldf}.
The EoS prior process includes EoS drawn from multiple underlying GPs with different parameters: $\log \sigma \in U(1,10)$  and $\ell\in U(0.1,0.9)$.
However, our GPs' kernels contain additional terms as well.
See Ref.~\cite{Essick:2019ldf} for more details.
In total we use $\sim2\times10^6$ draws, of which $\sim 3\times 10^5$ contribute to the prior nontrivially.

In Ref.~\cite{Miller:2021qha}, a single GP is used to generate EoS realizations using the same method.
This single GP is more tightly bound to the mean realization and nuclear models than corresponding piecewise-polytrope and spectral models due to the values of $\sigma=1$ and $\ell=1$ chosen.
Such values correspond to stronger correlations and over larger length scales than any GP we employ here.
This demonstrates that although the GP model is flexible, it is not necessarily agnostic.
This can be useful, for example, to examine the validity of a set of related nuclear models given astrophysical data~\cite{Essick:2020flb}.    

Figure~\ref{fig:GP_example_draws} shows example draws from our prior process plotted on top of posteriors for various parameters.  The candidate EoS exhibit a wide range of behavior as is perhaps most evident in the bottom panel.

\begin{table}
\begin{tabular}{ccc}
\hline
\hline
EoS prior process & Parameter & Prior  \\
\hline
\multirow{4}{*}{Spectral}
                   &       $r_0$        &             U(-4.37722, 4.91227)      \\
                   &       $r_1$        &                U(-1.82240, 2.06387)     \\
                   &       $r_2$        &             U(-0.32445, 0.36469)        \\
                   &       $r_3$        &            U(-0.09529, 0.11046)         \\
\hline
\multirow{4}{*}{Piecewise-polytrope}
                  &       $\log p_1$        &          U(33.6, 35.4)         \\
                  &      $\Gamma_1$        &               U(1.9, 4.5)    \\
                  &      $\Gamma_2$        &                U(1.1, 4.5)   \\
                  &      $\Gamma_3$        &               U(1.1, 4.5)\\
\hline
\multirow{5}{*}{Speed-of-sound}
                  &       $a_1$       & U(0.5, 1.5) \\
                  &       $a_2$       & U(1.3, 5) \\
                  &       $a_3$       & U(0.05, 3)  \\
                  &       $a_4$       &  U(1.5, 21)\\
                  &       $a_5$       & U(0.1, 1) \\
\hline
\end{tabular}
    \caption{
        List of parameters and corresponding priors on which each parametric EoS prior process depends.
    }
    \label{tab:parameterpriors}
\end{table}
% \begin{itemize}
% \end{itemize}
% \reed{
% There are additional hyperparam cuts described in appendix A of https://arxiv.org/abs/1910.09740, but the basic ranges are described in 
% \begin{itemize}
%     \item \url{https://ldas-jobs.ligo-wa.caltech.edu/~reed.essick/BNS_tides/gpr-eos-gw170817/eos/include/hadronic/gpr-agnostic}
%     \item \url{https://ldas-jobs.ligo-wa.caltech.edu/~reed.essick/BNS_tides/gpr-eos-gw170817/eos/include/hyperonic/gpr-agnostic}
%     \item \url{https://ldas-jobs.ligo-wa.caltech.edu/~reed.essick/BNS_tides/gpr-eos-gw170817/eos/include/quark/gpr-agnostic}
% \end{itemize}
% look for \texttt{min\_sigma} and the like.
% The exact ranges wiggle between compositions, but they're all pretty similar (smaller length scales than Miller+ and larger sigmas)
% }

%%%%%%%%%%%%%%%%%----------------------------------------------------------------------
\section{Description of the parametric EoS models}
\label{parametric-eos-appendix}

%-----------------------------------------------------------------------------------------
\subsection{Piecewise-polytrope parametrization}
\label{piecewise-polytrope-appendix}

In the piecewise-polytrope approach, consistent with Refs.~\cite{Read:2008iy,Carney:2018sdv}, the polytropic exponent is a piecewise constant function, which changes value at two predetermined densities
\begin{equation} 
    p(\rho) = \begin{cases}
        K_1 \rho^{\Gamma_1} : \rho < \rho_1 \\
        K_2 \rho^{\Gamma_2} : \rho_1 < \rho < \rho_2 \\
        K_3 \rho^{\Gamma_3} : \rho_2 < \rho
    \end{cases}
\end{equation}
Here $\rho_1 = 10^{14.7} \mathrm{g}/\mathrm{cm}^3$ and $\rho_2 = 10^{15} \mathrm{g}/\mathrm{cm}^3$ are fixed via an optimization for a set of candidate EoS following from nuclear models~\cite{Read:2008iy}.
The parameter $K_1$ is chosen to give some value $p_1 \equiv p(\rho_1)$, and $K_2$ and $K_3$ are then fixed by continuity. 
Therefore $\{\Gamma_1, \Gamma_2, \Gamma_3, p_1\}$ are the parameters in this model.
Their corresponding priors are given in Table~\ref{tab:parameterpriors}.
Extensions to this model with more polytropic segments or allowing the transition densities to vary are proposed in Refs.~\cite{Steiner:2010fz,Steiner:2015aea,Raithel:2016bux,OBoyle:2020qvf}.

 When $\{\Gamma_1, \Gamma_2, \Gamma_3, p_1\}$ are sampled from a uniform distribution then the resulting total EoS will be neither necessarily causal or stable. %: the EoS need not be neither monotonic nor have derivative less than $c^2$.
 Therefore, we have to enforce these constraints after the fact; specifically, we sample a set of parameters, compute the corresponding EoS, and save it only if it obeys causality and stability.\footnote{In practice, we impose a weaker causality constraint ($c_s \leq 1.1 c$) for our parametric models.}
 Overall, we retain \ppsamps EoS.
 We verified that this number is enough to efficiently characterize the posterior by confirming that we get consistent results with half as many draws.
 For the computation of the $p(\rho)$, and $\varepsilon(p)$ relations, we used LAL\textsc{Simulation}, a subsection of LAL\textsc{Suite}~\cite{lalsuite}.
 For checks of NS properties such as causality, we used LAL\textsc{Inference}~\cite{lalsuite, Veitch:2014wba}.
 Our priors are slightly more restrictive than those used in Ref.~\cite{Lackey:2014fwa} due to computational problems that arise for candidates with the highest $\Gamma_2, \Gamma_3$, which tend to represent acausal EoS candidates anyway.

%-----------------------------------------------------------------------------------------
\subsection{Spectral parametrization}
\label{spectral-appendix}

In the spectral approach, the polytropic exponent is expanded in a series of basis functions.
Following the conventions of Ref.~\cite{Lindblom:2010bb} which introduced the spectral parameterization, we take $x \equiv p/p_0$ where $p_0$ is the smallest pressure where the spectral parametrization will be used; the parametrization is matched to some other EoS at this density which serves as the low-density crust~\cite{Gamba:2019kwu}.
Then we set 

\begin{equation}
  p(\rho) = \rho^{\Gamma(x)}
\end{equation}
with
\begin{equation}
    \Gamma(x) = \sum_{i=0}^n \gamma_i \quant{\log(x)}^i
\end{equation} 
In most of the literature, and for our purposes $n$ is set to 3. 
Note that the overall scaling of $p(\rho)$ is fixed by $\gamma_0$, and again we have four total parameters $\{\gamma_0, \gamma_1, \gamma_2, \gamma_3 \}$.
In practice sampling individual parameters is impractical because generic combinations of parameters produce unphysical EoS, even if the parameter ranges are chosen carefully.
Instead, following Ref.~\cite{Wysocki:2020myz}, we sample in a different parameter space $r = (r_0, r_1, r_2, r_3)$ and under an affine map construct samples in $\gamma$.
The prior on $r$ is given in Table~\ref{tab:parameterpriors}.
Our analysis uses a total of \spsamps draws from the spectral model.
We again use the LAL\textsc{Suite} components LAL\textsc{Simulation} and LAL\textsc{Inference}~\cite{lalsuite, Veitch:2014wba}, with particular spectral components implemented by Ref.~\cite{Carney:2018sdv}.
The spectral EoS is stitched to a model of the SLy EoS just below $0.5\rhonuc$~\cite{Carney:2018sdv} (see Fig.~\ref{fig:process_prior_plot}).  

%-----------------------------------------------------------------------------------------
\subsection{Speed of sound parametrization}
\label{sos-appendix}

In this approach, the speed of sound is parametrized as a function of energy density.
Taking $z \equiv \veps/(\rho_{\nuc}c^2)$, we write
\begin{equation}
   \frac{c_s^2(z)}{c^2} = a_1 e^{-\half\quant{z-a_2}^2/a_3^2} + a_6 + \frac{\frac{1}{3}-a_6}{1+e^{-a_5\quant{z-a_4}}}
\end{equation}
with $a_1, a_2, a_3, a_4, a_5$ real parameters, and $a_6$ fixed by matching to a low-density crust.
In Ref.~\cite{Greif:2018njt}, the matching is done to a chiral effective field theory at $\sim \rhonuc$ with limits based on Fermi liquid theory enforced up to  a density of $1.5 \rhonuc$.
Since we do not wish to use more nuclear theory information for this model than others, we instead stitch to SLy at a density of $0.6 \rhonuc$, comparable to the stitching density of the spectral model.
Because of this, the parameter ranges in our implementation must be adjusted to generate realistic EoS candidates. The prior on each parameter is given in Table~\ref{tab:parameterpriors}.
Our analysis uses a total of \cssamps draws from this model. A similar model based on the speed of sound is presented in Ref.~\cite{Tews:2018kmu}.

%---------------------------------------------
\subsection{Causality in parametric models}

\begin{figure*}
    \centering
    \includegraphics[width=0.49\textwidth]{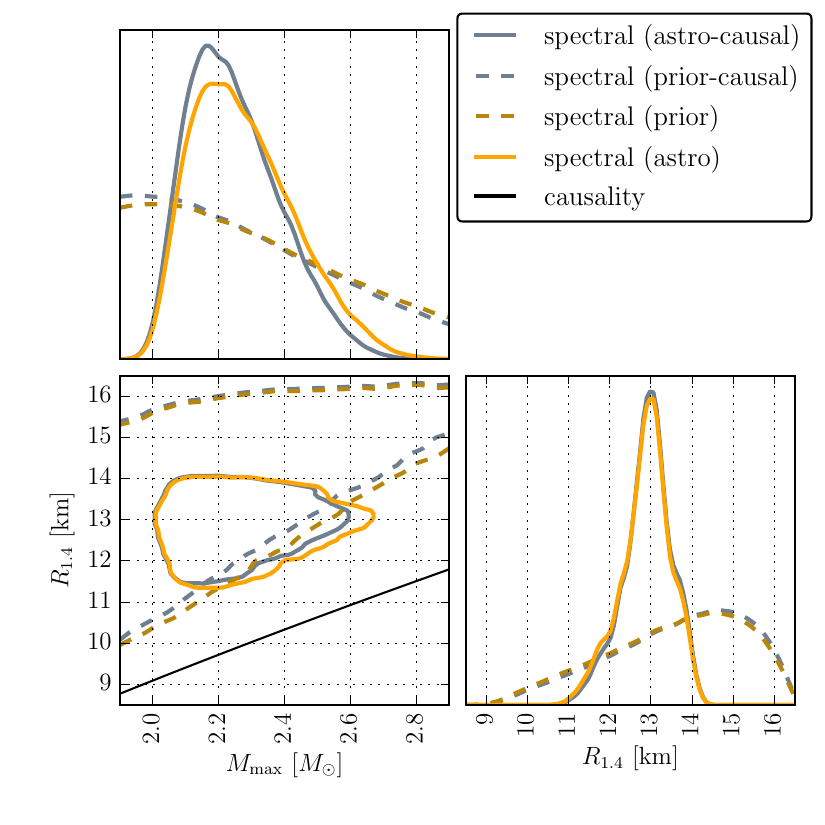}
    \includegraphics[width=0.49\textwidth]{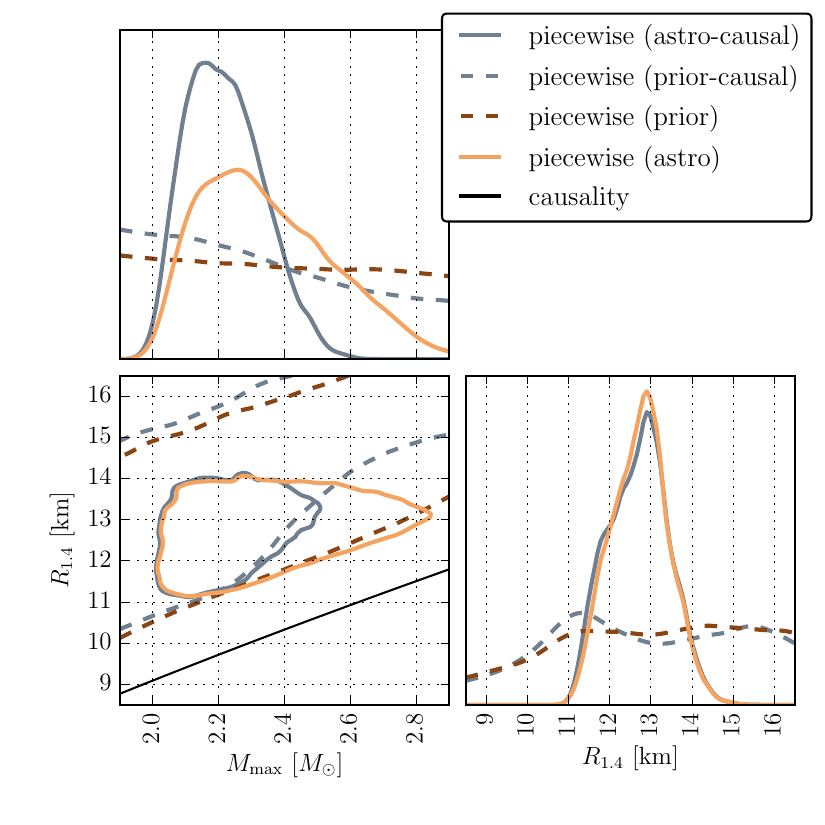} \\
    \includegraphics[width=0.49\textwidth]{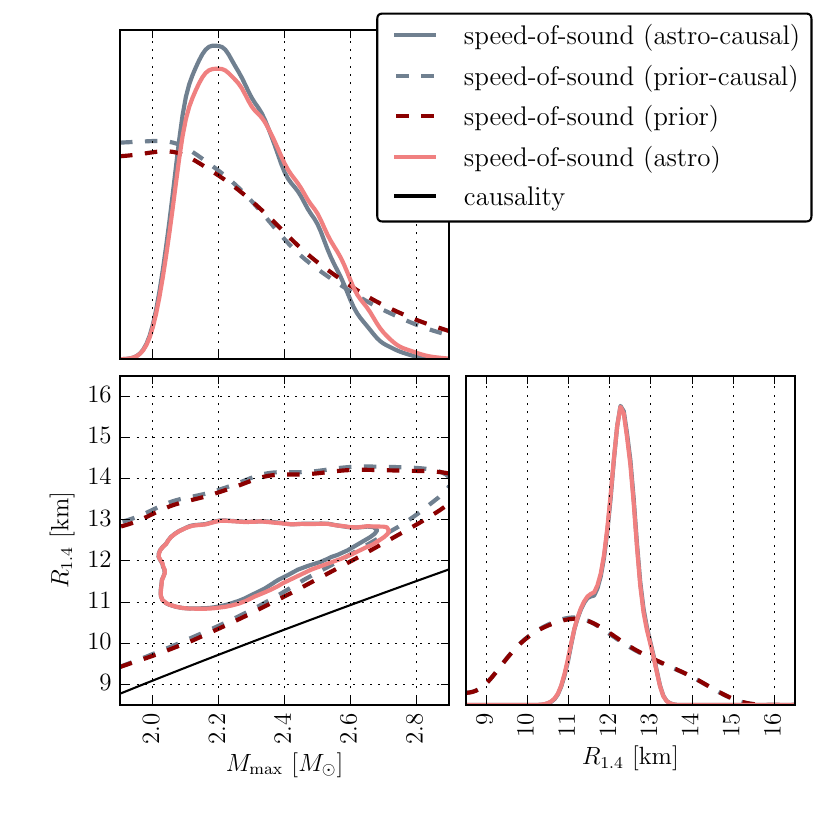}
    \caption{Comparison between using strictly causal ($c_s^2<c^2$) parametric EoS with each model (gray) and the headline results allowing some violation of the causal limit ($c_s^2<1.1c^2$).  When restricting to only causal EoS, the issues of model dependence and insufficient coverage of the physically allowed $\Mmax$-$\Rtyp$ space are more severe, especially for the piecewise-polytrope.}
    \label{fig:causality-enforced}
\end{figure*}

Because of the relatively large uncertainties, we follow Ref.~\cite{Carney:2018sdv} in not excluding parametric EoS until they have a large violation of the speed of sound $c_s > 1.1c$.
This is the standard criteria used in LAL\textsc{inference} for the piecewise-polytrope  and spectral models as part of determining if an EoS is physical.
For consistency we extend it to the speed-of-sound parametrization as well.
The primary motivation for this is to allow a possibly acausal EoS to represent another, causal EoS which is not modeled effectively by the prior on EoS~\cite{Carney:2018sdv}.
In addition, the LAL\textsc{inference} implementation of the spectral and piecewise-polytrope models enforces this criterion only up to the central density of the maximum mass NS.
In the speed-of-sound model, we require the EoS to be causal (or approximately causal) everywhere.
The nonparametric model obeys exact causality ($c_s \leq c$) at all densities. 

These choices were made for consistency with past work~\cite{Carney:2018sdv,Abbott:2018exr}, but we still find that this extra model freedom does not enable to spectral and piecewise-polytropic models to fill in the physically available $\Mmax$-$\Rtyp$ space up to the causality threshold.
Figure~\ref{fig:causality-enforced} shows that excluding the acausal models minimally affects the spectral and speed-of-sound results.
However, the piecewise-polytrope results are noticeably tighter, and the $\Mmax-\Rtyp$ allowed parameter space is covered significantly less. 
This also explains why the nominal piecewise-polytrope prior supports larger pressures than the nonparametric prior in Fig.~\ref{fig:prho_mr_all_models}.

%%%%%%%%%%%%%%%%%----------------------------------------------------------------------
\section{Further results with the parametric models}
\label{other-parametric-results}

\begin{figure*}
    \centering
    \includegraphics[width=.49\textwidth]{Figures/prho-sp_all-np-quantiles.pdf}
     \includegraphics[width=.49\textwidth]{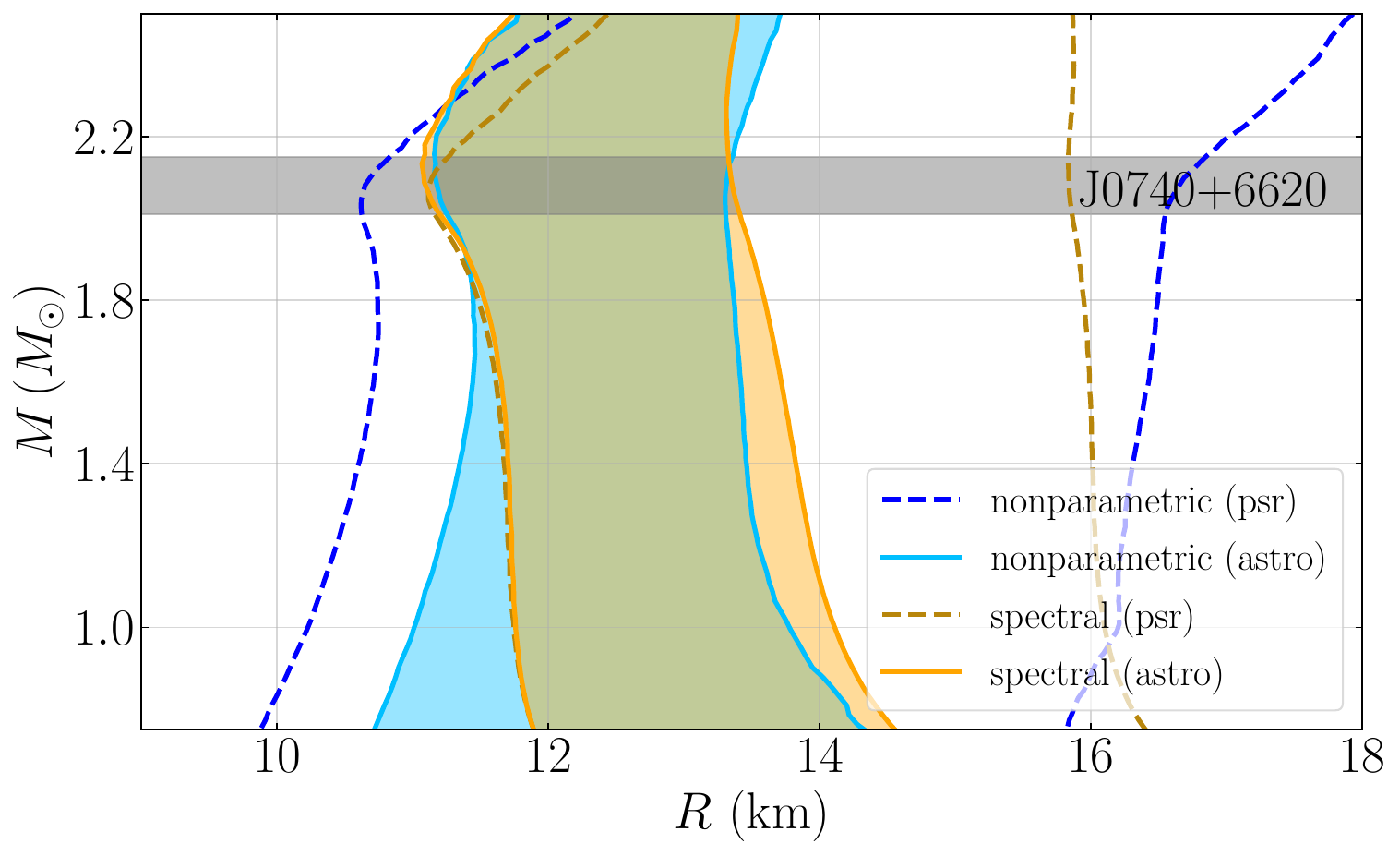}\\
    \includegraphics[width=.49\textwidth]{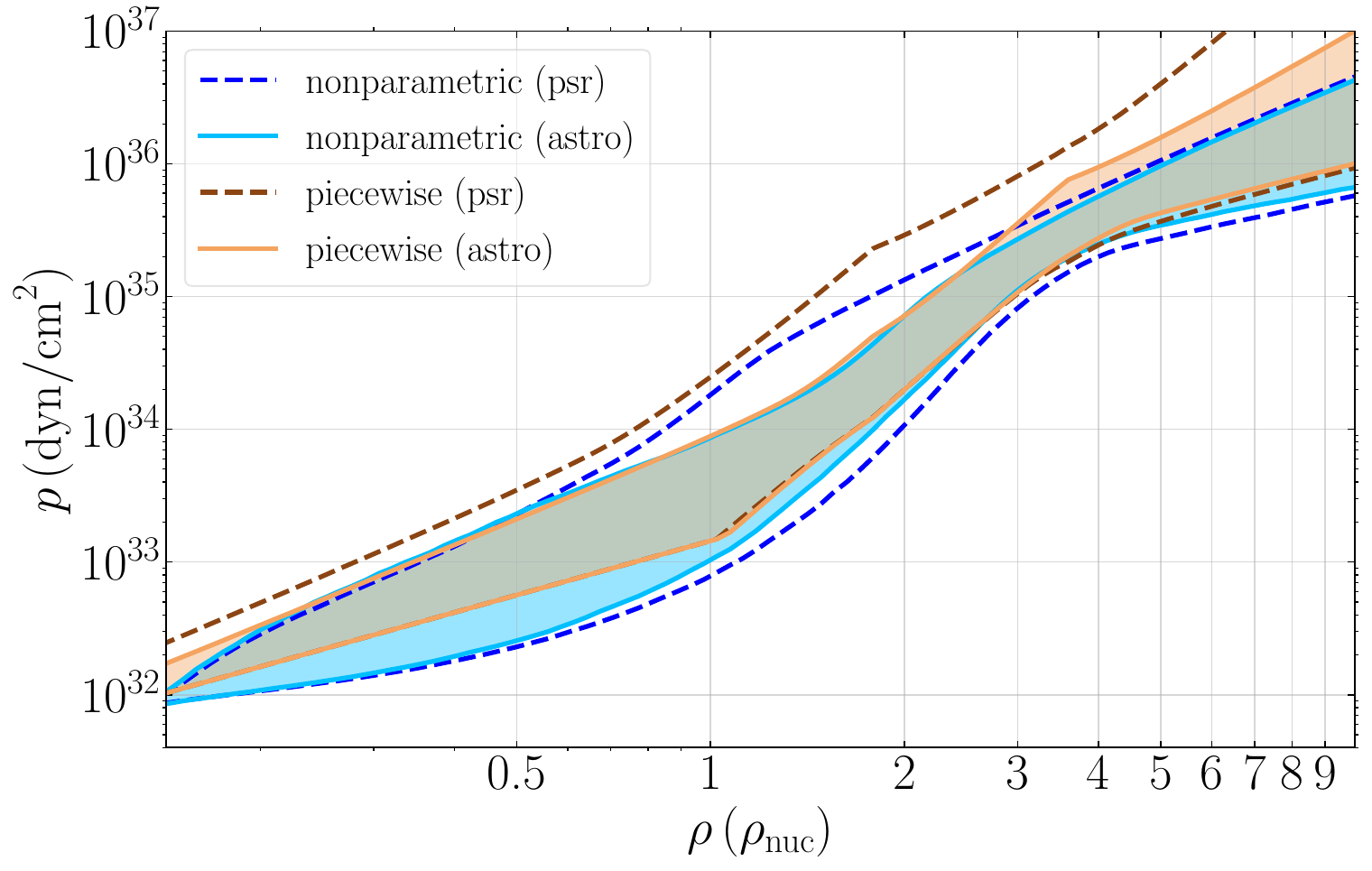}
    \includegraphics[width=.49\textwidth]{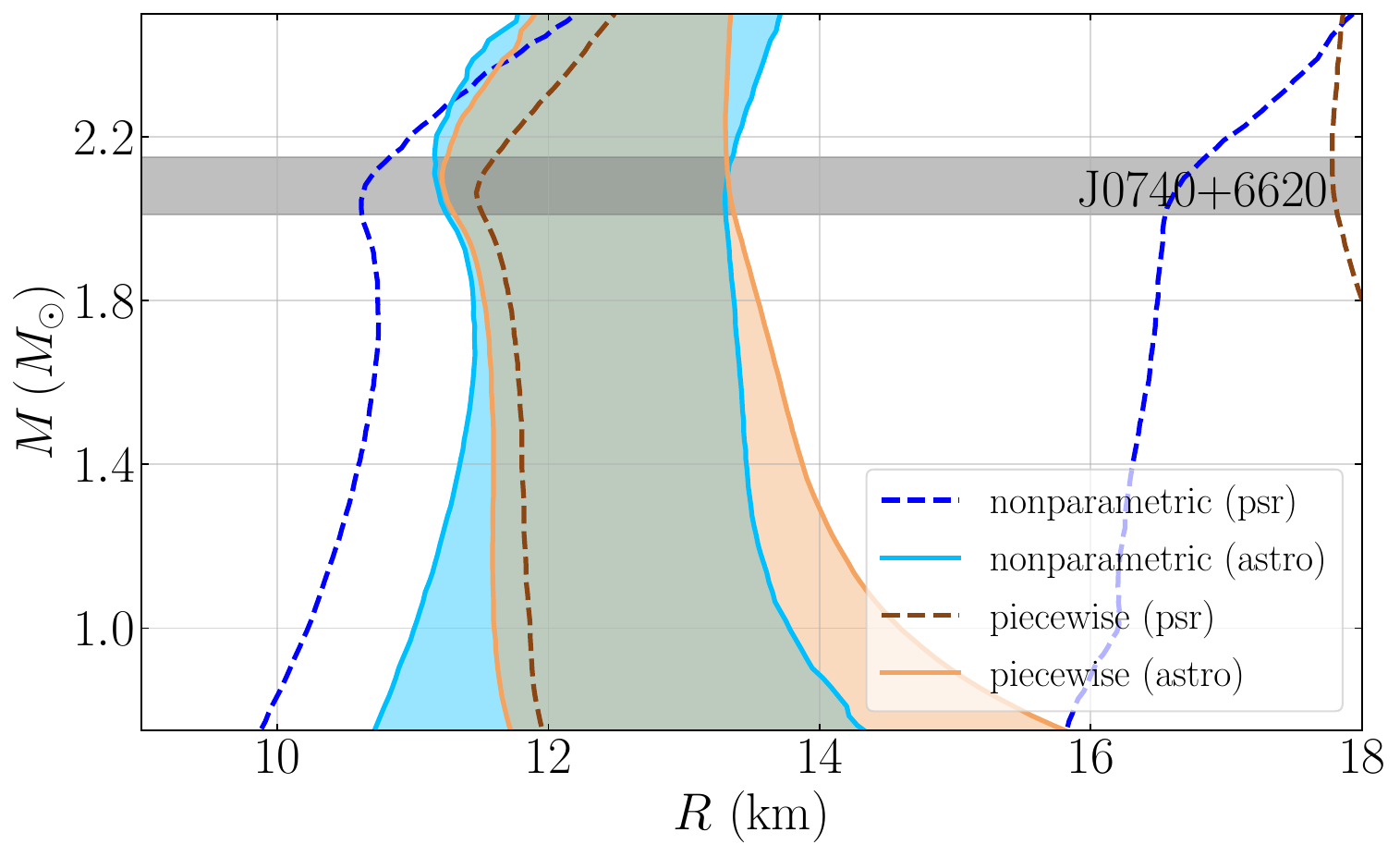}\\
     \includegraphics[width=.49\textwidth]{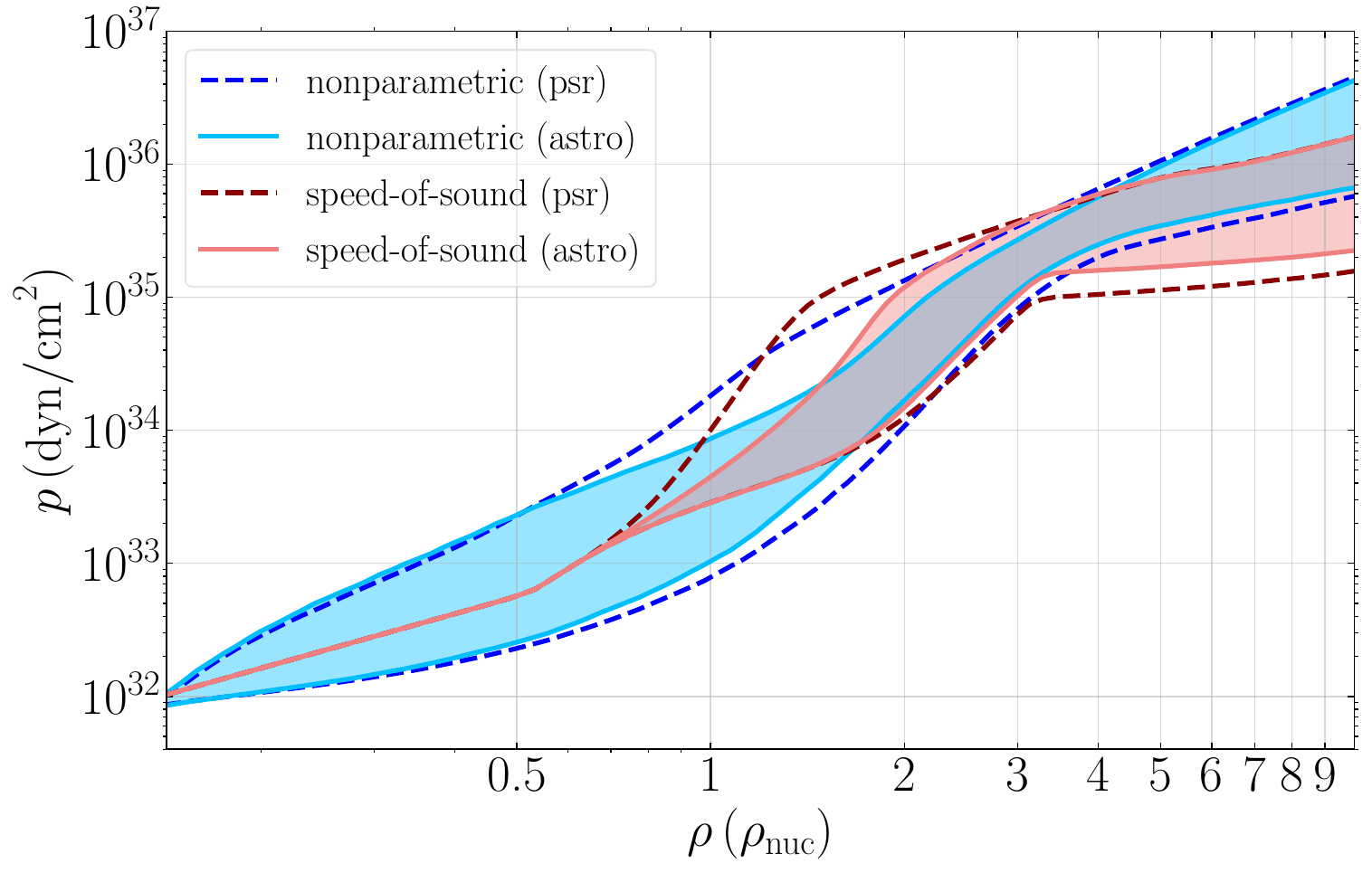}
    \includegraphics[width=.49\textwidth]{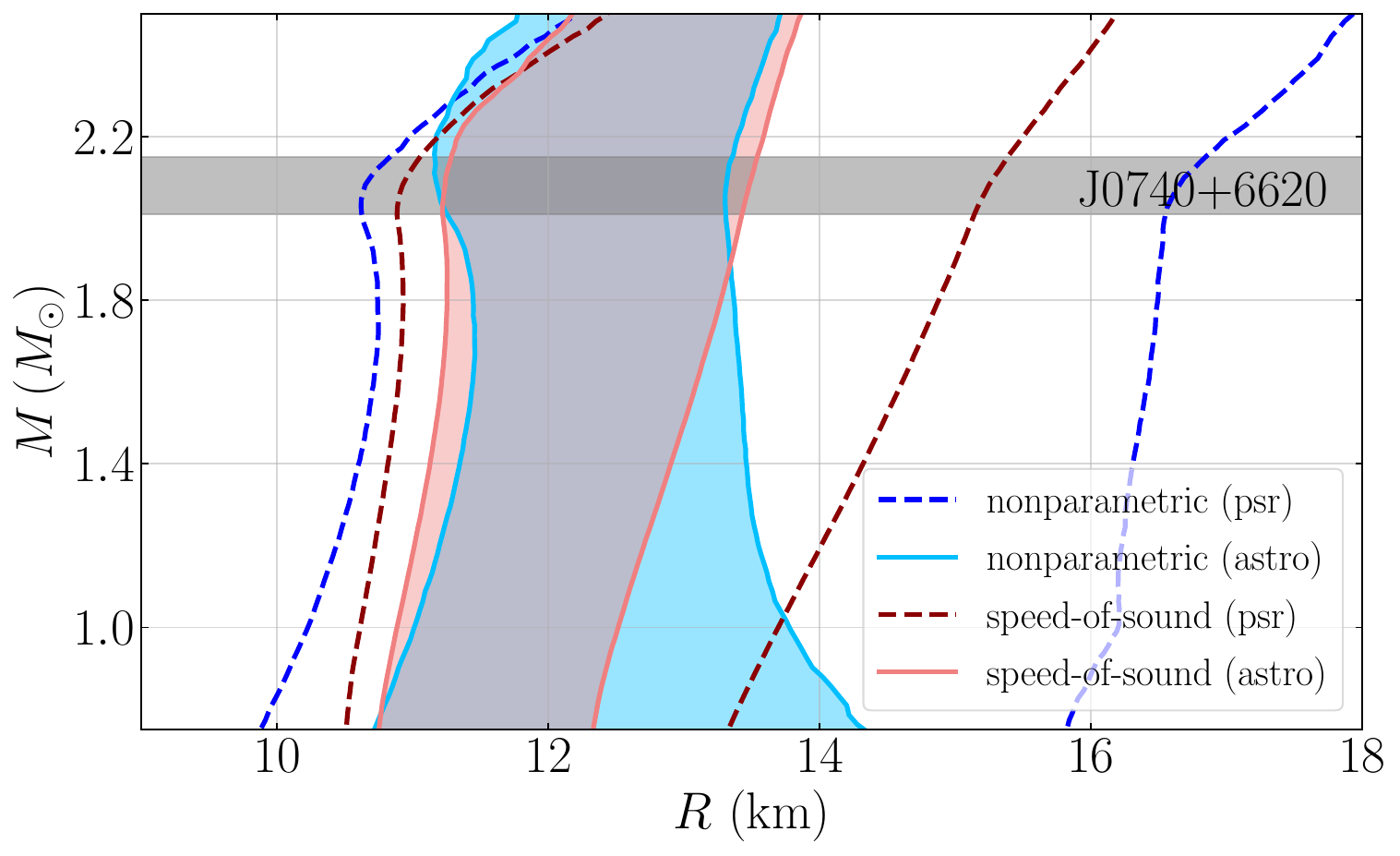}
    \caption{
        Symmetric 90\% credible region for the pressure $p$ at each density $\rho$ in units of the nuclear saturation density (left) and the radius $R$ as a function of the mass $M$.
        From top to bottom we show results with the spectral, piecewise-polytrope, and speed-of-sound parametric models.
        At each panel we overplot the corresponding nonparametric result for comparison.
        The spectral $p$-$\rho$ panel is identical to Fig.~\ref{fig:process_prior_plot}, but we show it for completeness.
        As in Fig.~\ref{fig:process_prior_plot} we show results with all astrophysical data (labeled ``astro," solid lines) and restricting to the heavy pulsars only (labeled ``psr," dashed lines).
    }
    \label{fig:prho_mr_all_models}
\end{figure*}

\begin{figure}
    \centering
    \includegraphics[width=.49\textwidth]{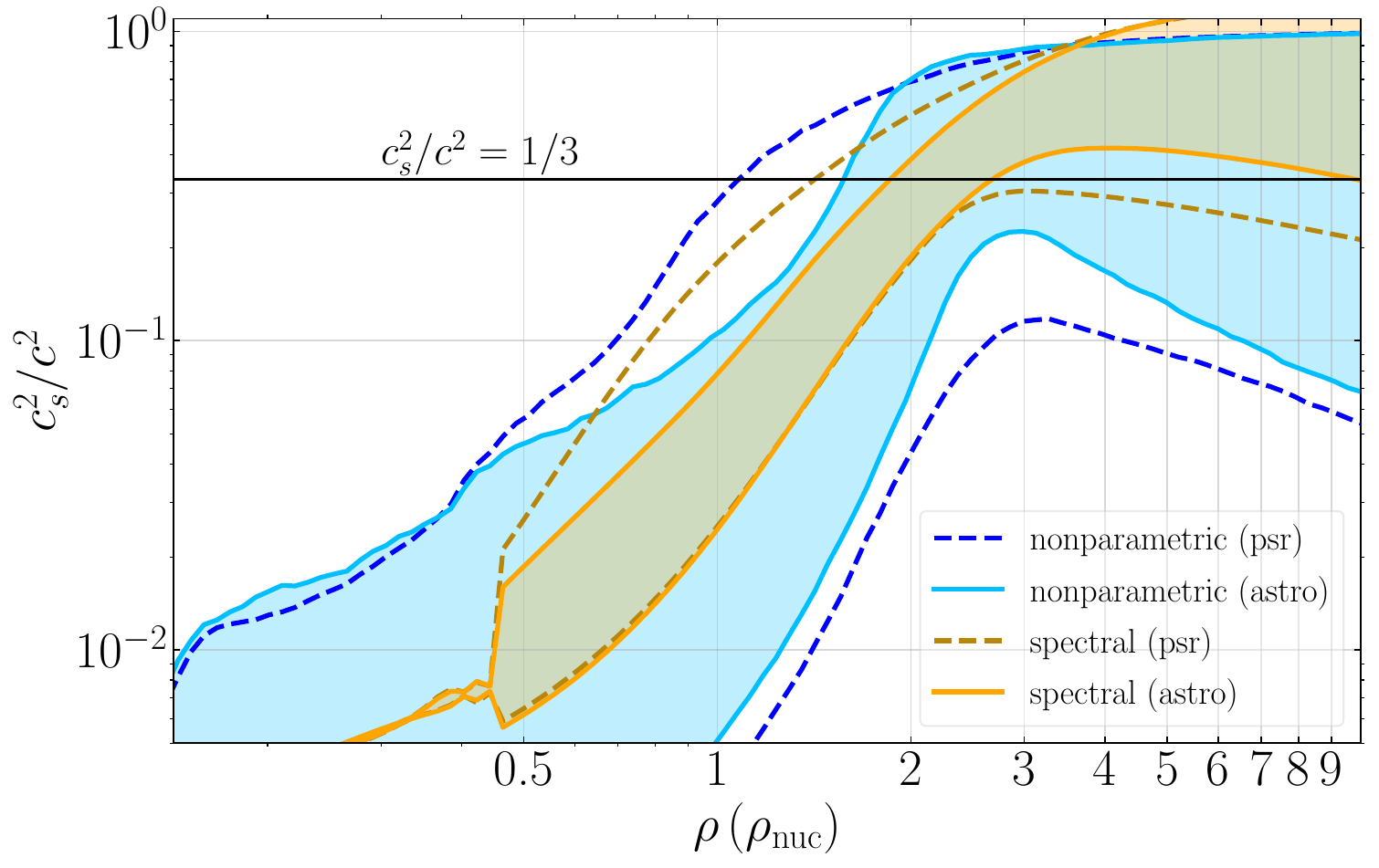}
    \includegraphics[width=.49\textwidth]{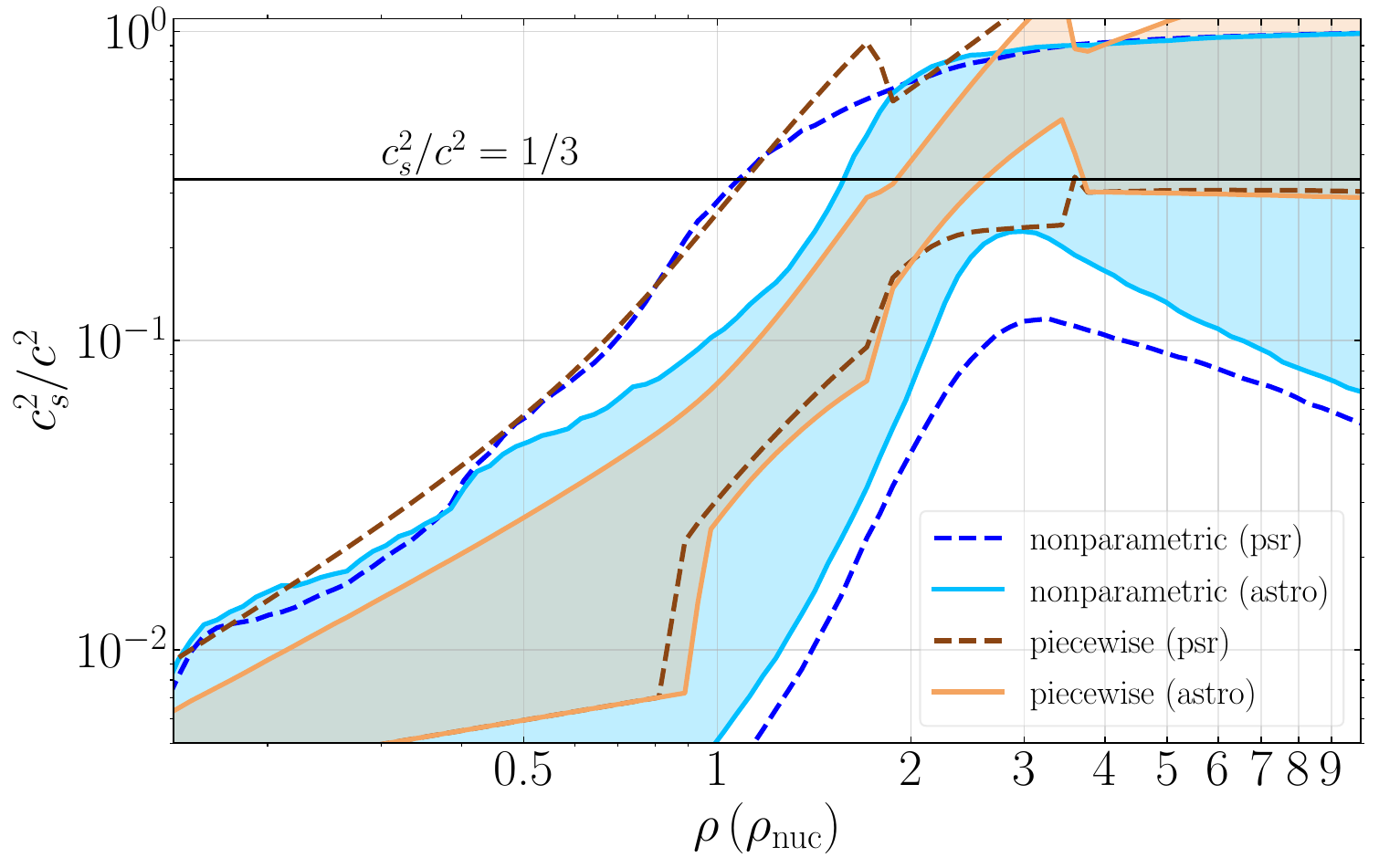}
    \includegraphics[width=.49\textwidth]{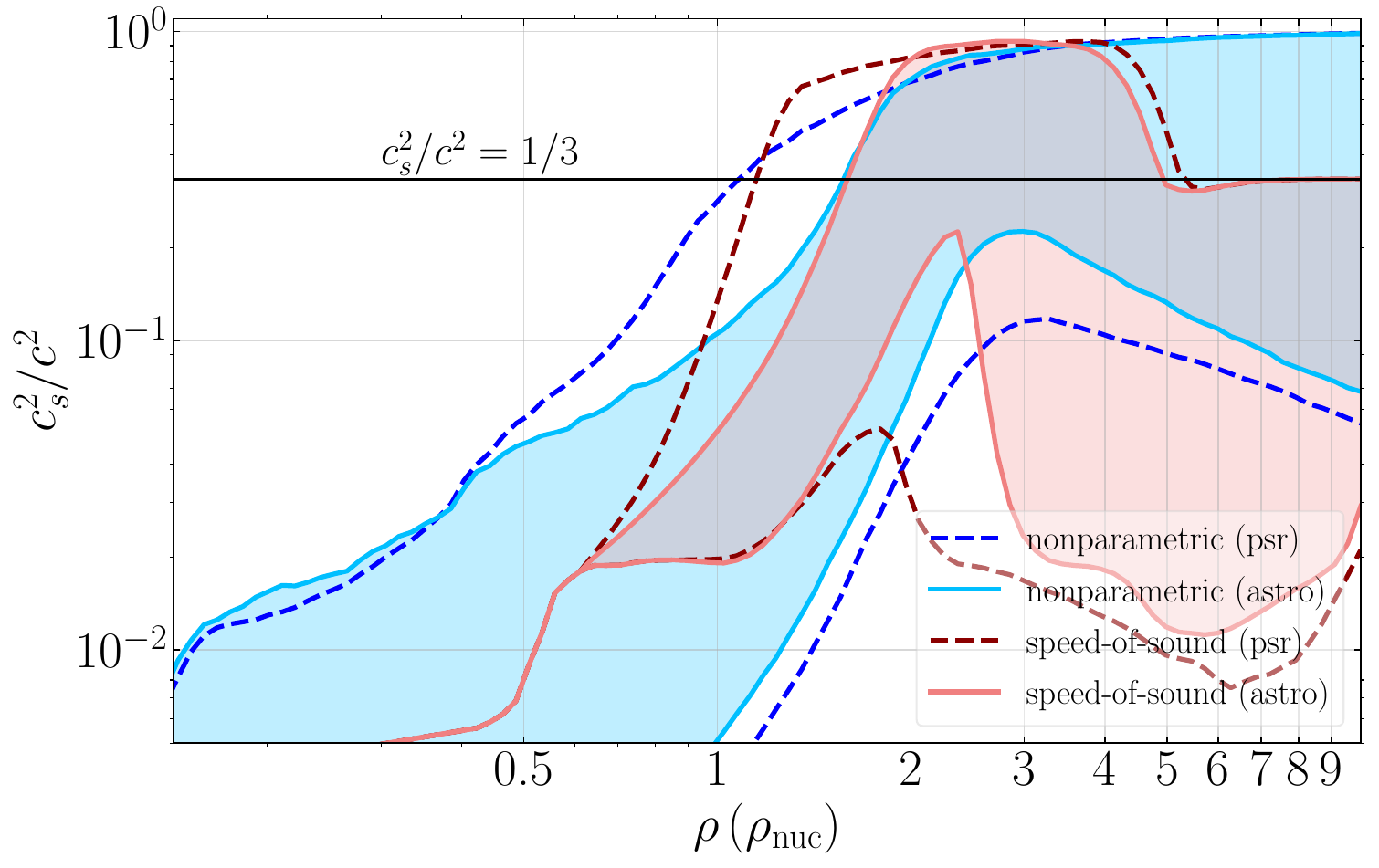}
    \caption{
        The speed of sound squared  as a function of baryon density in the spectral (top),  piecewise-polytrope (middle) models, and speed-of-sound (bottom) models, compared to our nonparametric model.
    }
    \label{fig:cs2_parametric_nonparametric}
\end{figure}

\begin{figure*}
    \centering
    \includegraphics[width=.49\textwidth]{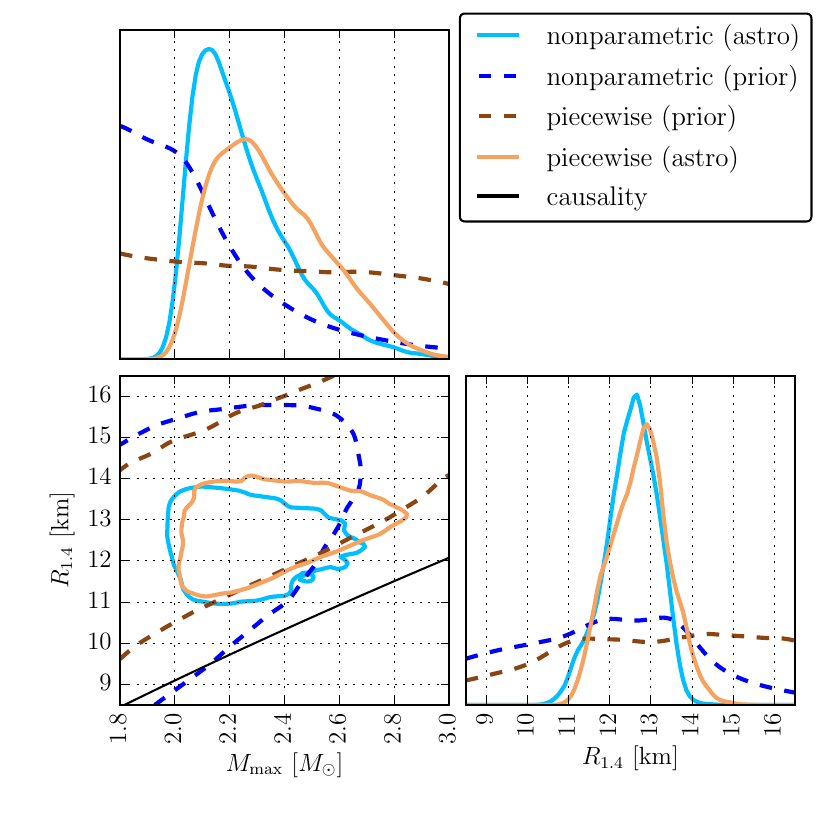}
    \includegraphics[width=.49\textwidth]{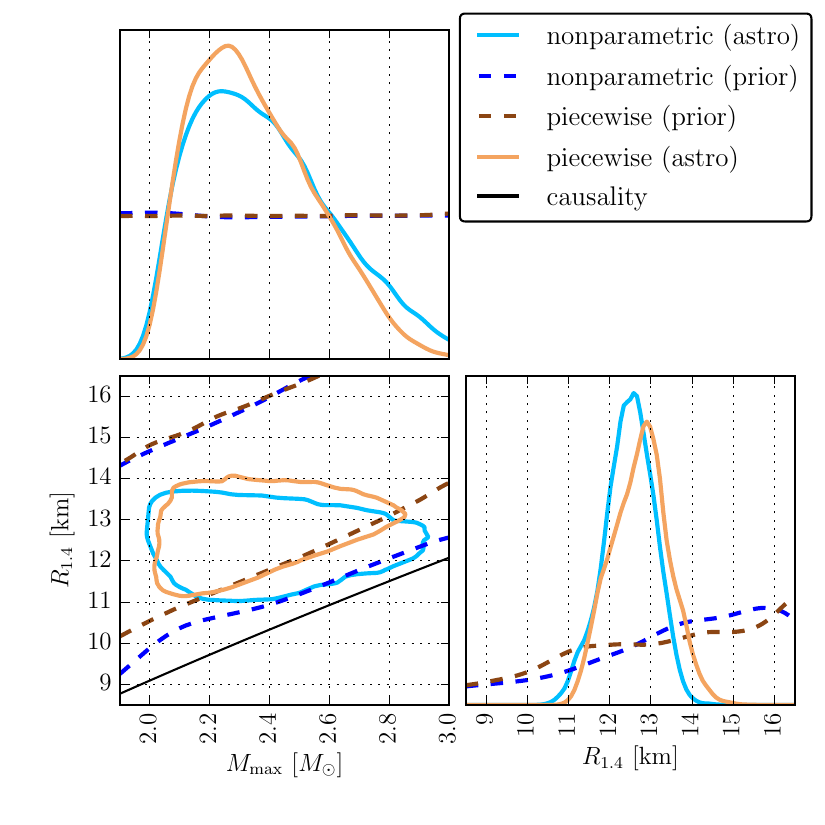}
    \includegraphics[width=.49\textwidth]{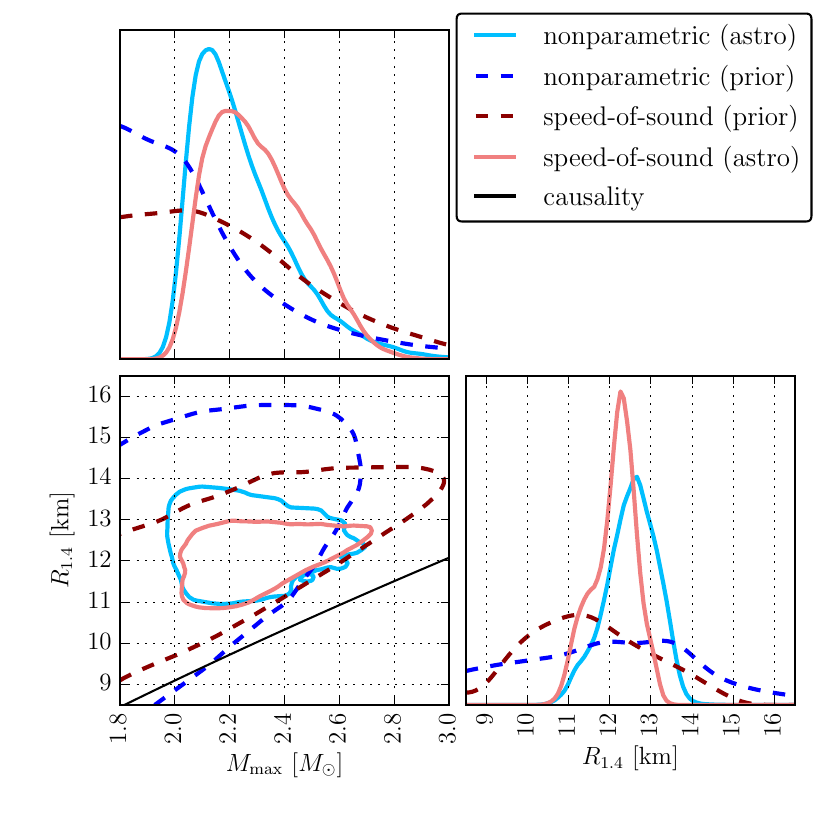}
    \includegraphics[width=.49\textwidth]{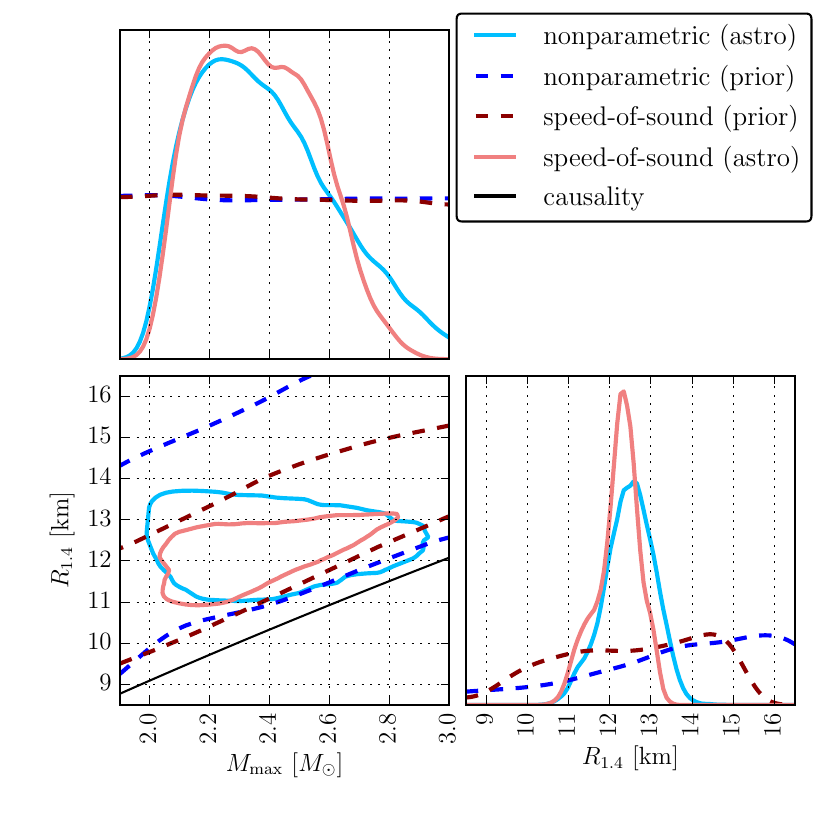}
    \caption{
        Similar to Fig.~\ref{fig:spectral_Mmax-Rtyp} but for the piecewise-polytrope (top panels) and the speed-of-sound (bottom panels) models.
        As with the spectral case, we find that the parametric model leads to tighter posteriors for $\Mmax$ compared to the nonparametric model.
        The two-dimensional plots show that this is again due to model-dependent correlations in $\Mmax$-$\Rtyp$.
    }
    \label{fig:piecewise-sos-MR}
\end{figure*}

Most results presented the main body of this paper were obtained using the spectral EoS model.
In this Appendix we present similar results with the piecewise-polytropic and the speed-of-sound models. Figure~\ref{fig:prho_mr_all_models} shows the pressure-density and mass-radius posteriors for all EoS models.
As expected, we find that the posteriors differ due to the different models.
Figure~\ref{fig:cs2_parametric_nonparametric} shows the posteriors for the speed of sound as a function of the density where again the nonparametric case results in the less constrained results as an outcome of larger model flexibility.

Figure~\ref{fig:piecewise-sos-MR} shows the equivalent of Fig.~\ref{fig:spectral_Mmax-Rtyp} for the piecewise-polytrope and the speed-of-sound models.
We again find that both parametric models lead to tighter constraints on $\Mmax$ for high values.
The two-dimensional plots show that model-dependent correlations between the maximum mass and the radius (equivalently low and high densities) exclude certain regions of the $\Mmax$-$\Rtyp$ space in the parametric marginal priors. % which are not excluded in the nonparametric case.
As with the spectral model, there is a gap between the piecewise-polytrope prior and the approximate causality threshold.  
The corresponding plots for the speed-of-sound parametrization show the opposite behavior: the prior reaches the causality threshold, but it fails to produce EoS with large $\Rtyp$ and small $\Mmax$. 

%%%%%%%%%%%%%%%%%%%%%%%%%%%%%%%%%%%%%%%%%%%%

\newpage
\bibliography{references}

%%%%%%%%%%%%%%%%%%%%%%%%%%%%%%%%%%%%%%%%%%%%
\end{document}

%% file: macros.tex
\newcommand{\spsamps}{\ensuremath{\sim 1.9\times 10^5}~}
\newcommand{\ppsamps}{\ensuremath{\sim 1.6\times 10^5}~}
\newcommand{\cssamps}{\ensuremath{\sim 1.6\times 10^5}~}